\documentclass[11pt,nofootinbib,showpacs]{revtex4}
\usepackage{amsfonts,amssymb,amsmath,graphicx,multirow,dsfont}
\def\dac{\displaystyle\frac}

\def\[{\left[}
\def\]{\right]}
\def\({\left(}
\def\){\right)}

\newcommand{\diag}{\mathop{\rm diag}\nolimits}

\begin{document}

\baselineskip7mm
\title{Cosmological dynamics of spatially flat Einstein-Gauss-Bonnet models in various dimensions: Low-dimensional $\Lambda$-term case}

\author{Sergey A. Pavluchenko}
\affiliation{Programa de P\'os-Gradua\c{c}\~ao em F\'isica, Universidade Federal do Maranh\~ao (UFMA), 65085-580, S\~ao Lu\'is, Maranh\~ao, Brazil}

\begin{abstract}
In this paper we perform a systematic study of spatially flat [(3+D)+1]-dimensional Einstein-Gauss-Bonnet cosmological models with $\Lambda$-term. We consider models that topologically are the product of 
two flat isotropic subspaces with different scale factors. One of these subspaces is three-dimensional and represents our space and the other is D-dimensional and represents extra dimensions. We consider 
no {\it Ansatz} on the 
scale factors, which makes our results quite general. With both Einstein-Hilbert and Gauss-Bonnet contributions in play, the cases with $D=1$ and $D=2$ have different dynamics due to the different 
structure of the equations of motion. We analytically study equations of motion in both cases and describe all possible regimes. It is demonstrated that $D=1$ case does not have physically viable regimes 
while $D=2$ has smooth transition from high-energy Kasner to anisotropic exponential regime. This transition occurs for two ranges of $\alpha$ and $\Lambda$:  $\alpha > 0$, $\Lambda > 0$ with 
$\alpha \Lambda \leqslant 1/2$ and $\alpha < 0$, $\Lambda > 0$ with $\alpha\Lambda < -3/2$. For the latter case, if $\alpha\Lambda = -3/2$, extra dimensional part has $h\to 0$ and so the size of extra
dimensions (in the sense of the scale factor) is reaching constant value.
We report substantial differences between $D=1$ and $D=2$ cases and between these cases and their vacuum counterparts,
describe features of the cases under study and discuss the origin of the differences.
\end{abstract}



\maketitle

\section{Introduction}

Einstein's general relativity was formulated more than hundred years ago, but the extra-dimensional models are even older. Indeed,
the first attempt to construct an extra-dimensional model was performed by Nordstr\"om~\cite{Nord1914} in 1914. It was a vector theory that unified Nordstr\"om's second gravity theory~\cite{Nord_2grav}
with Maxwell's electromagnetism. 
Later in 1915 Einstein introduced General Relativity (GR)~\cite{einst}, but still it took almost four years to prove that Nordstr\"om's theory and others were wrong. During the solar eclipse of 1919, the 
bending of light near the Sun was measured and the 
deflection angle was in perfect agreement with GR, while Nordstr\"om's theory, being scalar gravity, predicted a zeroth deflection angle.

But Nordstr\"om's idea about extra dimensions survived, and in 1919 Kaluza proposed~\cite{KK1} a similar model but based on GR: in his model five-dimensional Einstein equations could be decomposed into 
four-dimensional Einstein equations
plus Maxwell's electromagnetism. In order to perform such a decomposition, the extra dimensions should be ``curled'' or compactified into a circle and ``cylindrical conditions'' should be imposed. 
Later in 1926,
Klein proposed~\cite{KK2, KK3} a nice quantum mechanical interpretation of this extra dimension and so the theory called Kaluza-Klein was formally formulated. Back then their theory unified all
known interactions at that time. With time, more interactions were known and it became clear that to unify them all, more extra dimensions are needed. Nowadays, one of the promising theories to unify
all interactions is M/string theory.

Presence in the Lagrangian of the curvature-squared corrections is one of the distinguishing features of the  string theories gravitational counterpart.
Indeed, Scherk and Schwarz~\cite{sch-sch} were the
first to discover the potential presence of the $R^2$ and
$R_{\mu \nu} R^{\mu \nu}$ terms in the Lagrangian of the Virasoro-Shapiro
model~\cite{VSh1, VSh2}. A curvature squared term of the $R^{\mu \nu \lambda \rho}
R_{\mu \nu \lambda \rho}$ type appears~\cite{Candelas_etal} in the low-energy limit
of the $E_8 \times E_8$ heterotic superstring theory~\cite{Gross_etal} to match the kinetic term
for the Yang-Mills field. Later it was demonstrated~\cite{zwiebach} that the only
combination of quadratic terms that leads to a ghost-free nontrivial gravitation
interaction is the Gauss-Bonnet (GB) term:

$$
L_{GB} = L_2 = R_{\mu \nu \lambda \rho} R^{\mu \nu \lambda \rho} - 4 R_{\mu \nu} R^{\mu \nu} + R^2.
$$

\noindent This term, first found
by Lanczos~\cite{Lanczos1, Lanczos2} (therefore it is sometimes referred to
as the Lanczos term) is an Euler topological invariant in (3+1)-dimensional
space-time, but not in (4+1) and higher dimensions.
Zumino~\cite{zumino} extended Zwiebach's result on 
higher-than-squared curvature terms, supporting the idea that the low-energy limit of the unified
theory might have a Lagrangian density as a sum of contributions of different powers of curvature. In this regard the Einstein-Gauss-Bonnet (EGB) gravity could be seen as a subcase of more general Lovelock
gravity~\cite{Lovelock}, but in the current paper we restrain ourselves with only quadratic corrections and so to the EGB case.

Extra-dimensional theories have one thing in common---one needs to explain where additional dimensions are ``hiding'', since we do not sense them, at least with the current level of experiments. One of
the possible ways to hide extra dimensions, as well as to recover four-dimensional physics, is to build a so-called ``spontaneous compactification'' solution. Exact static solutions with the metric being a cross product of a
(3+1)-dimensional manifold and a constant curvature ``inner space''  were build for the first time in~\cite{add_1}, but with the (3+1)-dimensional manifold being Minkowski (the generalization for
a constant curvature Lorentzian manifold was done in~\cite{Deruelle2}).
In the context of cosmology, it is more interesting to consider a spontaneous compactification in the case where the four-dimensional part is given by a Friedmann-Robertson-Walker metric.
In this case it is also natural to consider the size of the extra dimensions as time dependent rather than static. Indeed in
\cite{add_4} it was explicitly shown  that in order to have a more realistic model one needs to consider the dynamical evolution of the extra-dimensional scale factor.
In~\cite{Deruelle2}, the equations of motion for compactification with both time-dependent scale factors were written for arbitrary Lovelock order in the special case of a spatially flat metric (the results were further proven in~\cite{prd09}).
The results of~\cite{Deruelle2} were reanalyzed for the special case of 10 space-time dimensions in~\cite{add_10}.
In~\cite{add_8}, the existence of dynamical compactification solutions was studied with the use of Hamiltonian formalism.
More recently, efforts to find spontaneous  compactifications were made in~\cite{add13}, where
the dynamical compactification of the (5+1) Einstein-Gauss-Bonnet model was considered; in \cite{MO04, MO14} with different metric {\it Ans\"atze} for scale factors
corresponding to (3+1)- and extra-dimensional parts; and in \cite{CGP1, CGP2, CGPT}, where general (e.g., without any {\it Ansatz}) scale factors and curved manifolds were considered. Also, apart from
cosmology, the recent analysis has focused on
properties of black holes in Gauss-Bonnet~\cite{alpha_12, add_rec_1, add_rec_2, addn_1, addn_2} and Lovelock~\cite{add_rec_3, add_rec_4, addn_3, addn_4} gravities, features of gravitational collapse in these
theories~\cite{addn_5, addn_6, addn_7}, general features of spherical-symmetric solutions~\cite{addn_8}, and many others.

In the context of finding exact solutions, the most common {\it Ansatz} used for the functional form of the scale factor is exponential or power law.
Exact solutions with exponential functions  for both  the (3+1)- and extra-dimensional scale factors were studied for the first time in  \cite{Is86}, and an exponentially increasing (3+1)-dimensional
scale factor and an exponentially shrinking extra-dimensional scale factor were described.
Power-law solutions have been analyzed  in \cite{Deruelle1, Deruelle2} and more  recently in~\cite{mpla09, prd09, Ivashchuk, prd10, grg10} so that there is an  almost complete description
(see also~\cite{PT} for useful comments regarding physical branches of the solutions).
Solutions with exponential scale factors~\cite{KPT} have been studied in detail, namely, models with both variable~\cite{CPT1} and constant~\cite{CST2} volume, developing a general scheme for
constructing solutions in EGB; recently~\cite{CPT3} this scheme
was generalized for general Lovelock gravity of any order and in any dimensions. Also, the stability of the solutions was addressed in~\cite{my15} (see also~\cite{iv16} for stability of general exponential
solutions in EGB gravity), where it was
demonstrated that only a handful of the solutions could be called ``stable'', while the remaining are either unstable or have neutral/marginal stability, and so additional investigation is
required.

In order to find all possible regimes of Einstein-Gauss-Bonnet cosmology, it is necessary to go beyond an exponential or power-law {\it Ansatz} and keep the functional form of the scale factor generic.
We are also particularly interested in models that allow dynamical compactification, so it is natural to
consider the metric as the product of a spatially three-dimensional part and an extra-dimensional part. In that case the three-dimensional part represents ``our Universe'' and we expect for this part to 
expand
while the extra-dimensional part should be suppressed in size with respect to the three-dimensional one. In \cite{CGP1} it was found that there exists a phenomenologically
sensible regime in the case when the curvature of the extra dimensions is negative and the Einstein-Gauss-Bonnet theory does not admit a maximally symmetric solution. In this case the
three-dimensional Hubble parameter and the extra-dimensional scale factor asymptotically tend to the constant values. In \cite{CGP2} a detailed analysis of the cosmological dynamics in this model
with generic couplings was performed. Recently this model was also studied in~\cite{CGPT}, where it was demonstrated that, with an additional constraint on couplings, Friedmann-type late-time behavior
could be restored.

The current paper is a spiritual successor of~\cite{my16a}, where we investigated cosmological dynamics of the vacuum Einstein-Gauss-Bonnet model. In both papers the spatial section is a product of two 
spatially flat
manifolds with one of them three-dimensional, which represents our Universe and the other is extra-dimensional. In~\cite{my16a} we considered vacuum model while in the current paper -- the model with the 
cosmological
term. In~\cite{my16a} we demonstrated that the vacuum model has two physically viable regimes -- first of them is the smooth transition from high-energy GB Kasner to low-energy GR Kasner. This regime
appears for $\alpha > 0$ at $D=1,\,2$ and for $\alpha < 0$ at $D \geqslant 2$ (so that at $D=2$ it appears for both signs of $\alpha$). The other viable regime is smooth transition from high-energy GB 
Kasner to anisotropic exponential regime with expanding three-dimensional section (``our Universe'') and contracting extra dimensions; this regime occurs only for $\alpha > 0$ and at $D \geqslant 2$.
Let us note that in~\cite{CGP1, CGP2, CGPT} we considered similar model but with both manifolds to be constant (generally non-zero) curvature. Unlike the paper with vacuum solutions, in this paper we
limit ourselves with only lower-dimensional ($D=1$ and $D=2$) cases; the higher-dimensional cases -- $D=3$ and the general $D\geqslant 4$ case -- will be considered in a separate forthcoming paper.

The structure of the manuscript is as follows: first we write down general equations of motion for Einstein-Gauss-Bonnet gravity, then we rewrite them for our symmetry {\it Ansatz}. In the following
sections we analyze them for $D=1$ and $D=2$ cases, considering the $\Lambda$-term case in this paper only. Each case is
followed by a small discussion of the results and properties of this particular case; after considering all cases we discuss their properties, generalities, and differences
and draw conclusions.

\section{Equations of motion}

As mentioned above, we consider the spatially flat anisotropic cosmological model in Einstein-Gauss-Bonnet gravity with $\Lambda$-term as a matter source.
The equations of motion for such model include both first and second Lovelock contributions and could easily be derived from the general case (see, e.g.,~\cite{prd09}). We consider flat
anisotropic metric

\begin{equation}\label{metric}
g_{\mu\nu} = \diag\{ -1, a_1^2(t), a_2^2(t),\ldots, a_N^2(t)\};
\end{equation}

\noindent the Lagrangian of this theory has the form

\begin{equation}\label{lagr}
{\cal L} = R + \alpha{\cal L}_2 - 2\Lambda,
\end{equation}

\noindent where $R$ is the Ricci scalar and ${\cal L}_2$,

\begin{equation}
{\cal L}_2 = R_{\mu \nu \alpha \beta} R^{\mu \nu \alpha \beta} - 4
R_{\mu \nu} R^{\mu \nu} + R^2 \label{lagr1}
\end{equation}

\noindent is the Gauss-Bonnet Lagrangian. Then substituting (\ref{metric}) into the Riemann and Ricci tensors and the scalar in (\ref{lagr}) and (\ref{lagr1}), and varying (\ref{lagr}) with respect to
the metric, we obtain the equations of motion,

\begin{equation}
\begin{array}{l}
2 \[ \sum\limits_{j\ne i} (\dot H_j + H_j^2)
+ \sum\limits_{\substack{\{ k > l\} \\ \ne i}} H_k H_l \] + 8\alpha \[ \sum\limits_{j\ne i} (\dot H_j + H_j^2) \sum\limits_{\substack{\{k>l\} \\ \ne \{i, j\}}} H_k H_l +
3 \sum\limits_{\substack{\{ k > l >  \\   m > n\} \ne i}} H_k H_l
H_m H_n \] - \Lambda = 0
\end{array} \label{dyn_gen}
\end{equation}

\noindent as the $i$th dynamical equation. The first Lovelock term---the Einstein-Hilbert contribution---is in the first set of brackets and the second term---Gauss-Bonnet---is in the second set; $\alpha$
is the coupling constant for the Gauss-Bonnet contribution and we put the corresponding constant for Einstein-Hilbert contribution to unity. Also, since we a consider spatially flat cosmological model, scale
factors do not hold much in the physical sense and the equations are rewritten in terms of the Hubble parameters $H_i = \dot a_i(t)/a_i(t)$. Apart from the dynamical equations, we write down a constraint equation

\begin{equation}
\begin{array}{l}
2 \sum\limits_{i > j} H_i H_j + 24\alpha \sum\limits_{i > j > k > l} H_i H_j H_k H_l = \Lambda.
\end{array} \label{con_gen}
\end{equation}

As mentioned in the Introduction,
we want to investigate the particular case with the scale factors split into two parts -- separately three dimensions (three-dimensional isotropic subspace), which are supposed to represent our world, and the remaining represent the extra dimensions ($D$-dimensional isotropic subspace). So we put $H_1 = H_2 = H_3 = H$ and $H_4 = \ldots = H_{D+3} = h$ ($D$ designs the number of additional dimensions) and the
equations take the following form: the
dynamical equation that corresponds to $H$,

\begin{equation}
\begin{array}{l}
2 \[ 2 \dot H + 3H^2 + D\dot h + \dac{D(D+1)}{2} h^2 + 2DHh\] + 8\alpha \[ 2\dot H \(DHh + \dac{D(D-1)}{2}h^2 \) + \right. \\ \\ \left. + D\dot h \(H^2 + 2(D-1)Hh + \dac{(D-1)(D-2)}{2}h^2 \) +
2DH^3h + \dac{D(5D-3)}{2} H^2h^2 + \right. \\ \\ \left. + D^2(D-1) Hh^3 + \dac{(D+1)D(D-1)(D-2)}{8} h^4 \] - \Lambda=0,
\end{array} \label{H_gen}
\end{equation}

\noindent the dynamical equation that corresponds to $h$,

\begin{equation}
\begin{array}{l}
2 \[ 3 \dot H + 6H^2 + (D-1)\dot h + \dac{D(D-1)}{2} h^2 + 3(D-1)Hh\] + 8\alpha \[ 3\dot H \(H^2 + 2(D-1)Hh + \right .\right. \\ \\ \left. \left. + \dac{(D-1)(D-2)}{2}h^2 \) +  (D-1)\dot h \(3H^2 + 3(D-2)Hh +
\dac{(D-2)(D-3)}{2}h^2 \) + 3H^4 +
\right. \\ \\ \left. + 9(D-1)H^3h + 3(D-1)(2D-3) H^2h^2 +  \dac{3(D-1)^2 (D-2)}{2} Hh^3 + \right. \\ \\ \left. + \dac{D(D-1)(D-2)(D-3)}{8} h^4 \] - \Lambda =0,
\end{array} \label{h_gen}
\end{equation}

\noindent and the constraint equation,

\begin{equation}
\begin{array}{l}
2 \[ 3H^2 + 3DHh + \dac{D(D-1)}{2} h^2 \] + 24\alpha \[ DH^3h + \dac{3D(D-1)}{2}H^2h^2 + \dac{D(D-1)(D-2)}{2}Hh^3 + \right. \\ \\ \left. + \dac{D(D-1)(D-2)(D-3)}{24}h^4\] = \Lambda.
\end{array} \label{con2_gen}
\end{equation}

Looking at (\ref{H_gen}) and (\ref{h_gen}) one can see that for $D\geqslant 4$ the equations of motion contain the same terms, while for $D=\{1, 2, 3\}$ the terms are different [say, for $D=3$ terms with
the $(D-3)$ multiplier are absent and so on] and the dynamics should be different also.
We are going to study these four cases separately. As we mentioned in the Introduction, in this paper we are going to consider only the $\Lambda$-term case; the vacuum case we considered in the previous
paper~\cite{my16a}
while the general case with a perfect fluid with an arbitrary equation of state we as well as the effect of curvature, are going to be considered in the papers to follow. As we also noted in the Introduction, in this
particular paper we consider only $D=\{1, 2\}$ cases -- the $D=3$ and general $D\geqslant 4$ cases we consider in forthcoming paper.

\section{$D=1$ case}

In this case the equations of motion take the form ($H$-equation, $h$-equation, and constraint correspondingly)

\begin{equation}
\begin{array}{l}
4\dot H + 6H^2 + 2\dot h + 2h^2 + 4Hh + 8\alpha \( 2(\dot H + H^2)Hh + (\dot h + h^2)H^2\) = \Lambda,
\end{array} \label{D1_H}
\end{equation}

\begin{equation}
\begin{array}{l}
6\dot H + 12H^2 + 24\alpha (\dot H + H^2)H^2 = \Lambda,
\end{array} \label{D1_h}
\end{equation}

\begin{equation}
\begin{array}{l}
6H^2 + 6Hh + 24\alpha H^3h = \Lambda.
\end{array} \label{D1_con}
\end{equation}

\begin{figure}
\includegraphics[width=1.0\textwidth, angle=0]{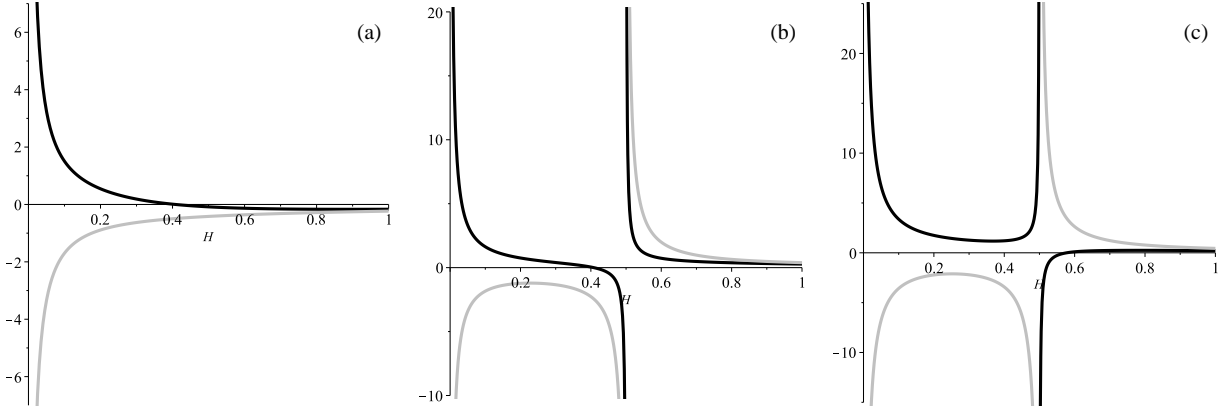}
\caption{Behavior of $h(H)$ from Eq. (\ref{D1_hh}) in $D=1$ case. Panel (a) corresponds to $\alpha > 0$ while both (b) and (c) -- to $\alpha < 0$. In all panels black curve corresponds to the typical $\Lambda > 0$ behavior while grey -- to $\Lambda < 0$. The difference between (b) and (c) panels is that for black curves in (b) we have $\alpha\Lambda > -3/2$ while in (c)
$\alpha\Lambda < -3/2$
(see the text for more details).}\label{D1_1}
\end{figure}

From (\ref{D1_con}) we can easily see that

\begin{equation}
\begin{array}{l}
h =  \dac{\Lambda - 6H^2}{6H(1+4\alpha H^2)},
\end{array} \label{D1_hh}
\end{equation}

\noindent and we present them in Fig.~\ref{D1_1}. In there, panel (a) corresponds to $\alpha > 0$ while both (b) and (c) -- to $\alpha < 0$. In all panels black curve corresponds to the typical $\Lambda > 0$ behavior while grey -- to $\Lambda < 0$. The difference between (b) and (c) panels is that for black curves in (b) we have $\alpha\Lambda > -3/2$ and in (c) $\alpha\Lambda <-3/2$.
The difference in behavior between $\alpha\Lambda > -3/2$ and $\alpha\Lambda < -3/2$ cases manifests itself only for $\alpha < 0$, that is why we have only one curve for $\alpha > 0$. Also, one can see that grey curves in Figs.~\ref{D1_hh}(b) and (c) coincide -- they both are for $\alpha < 0$ and $\Lambda < 0$ so for both of them $\alpha\Lambda > -3/2$. Let us also note that for successful compactification one needs $H>0$ and $h\leqslant 0$ so that from Fig.~\ref{D1_1} one can judge about the
regions of the initial conditions and parameters where it could be satisfied.

Now with explicit for $h(H)$ we can solve (\ref{D1_H})--(\ref{D1_h}) and substitute (\ref{D1_hh}) into them to get

\begin{equation}
\begin{array}{l}
\dot H = \dac{\Lambda - 24\alpha H^4 - 12H^2}{6(1+4\alpha H^2)}, \\ \\
\dot h = - \dac{ 576\alpha^2H^8 - 288\alpha^2\Lambda H^6 + 144\alpha H^6 - 192\alpha\Lambda H^4 + 12\alpha \Lambda^2 H^2 - 72 H^4 - 6\Lambda H^2 + \Lambda^2 }{36H^2 (1+4\alpha H^2)^3}.
\end{array} \label{D1_dHdh}
\end{equation}

Before considering $\dot H(H)$ and $\dot h(H)$ curves, let us analyze (\ref{D1_dHdh}), finding their roots and asymptotes. As $\dot H(H)$ and $\dot h(H)$ have the same denominator, they have the same asymptote
located at $(1+4\alpha H^2) = 0$, which corresponds to $H^2 = -1/(4\alpha)$, which is the same asymptote as from (\ref{D1_hh}), so that $H^2 = -1/(4\alpha)$ is nonstandard singularity\footnote{Nonstandard singularity is the
situation when the highest derivative ($\dot H$ and/or $\dot h$ in our case) diverge while lower derivatives and/or variables are regular. This situation is singular, but due to regularity of lower derivatives and/or variables,
it happens some finite time. We discuss this situation more in the Discussions section.}
in this case. As of the roots, they are the following:

\begin{equation}
\begin{array}{l}
\dot H = 0 ~ \Leftrightarrow ~ H^2_\pm = - \dac{3\pm\sqrt{6\alpha\Lambda + 9}}{12\alpha}, \\ \\
\dot h = 0 ~ \Leftrightarrow ~ H^2 = - \dac{3\pm\sqrt{6\alpha\Lambda + 9}}{12\alpha}, ~\dac{6\alpha\Lambda + 3 \pm\sqrt{36\alpha^2\Lambda^2 + 60\alpha\Lambda + 9}}{24\alpha}.
\end{array} \label{D1_dHdh_0}
\end{equation}

Before going further, let us note an interesting fact -- roots of $\dot H = 0$ are the roots of $\dot h = 0$ as well, so they determine exponential solutions allowed in this case. If we substitute corresponding
$H$ from (\ref{D1_dHdh_0}) into (\ref{D1_hh}), we can note that both of them have $h=H$. So that in $D=1$ case we have two exponential solutions and both of them are isotropic -- which is very different
from the vacuum case.

\begin{figure}
\includegraphics[width=1.0\textwidth, angle=0]{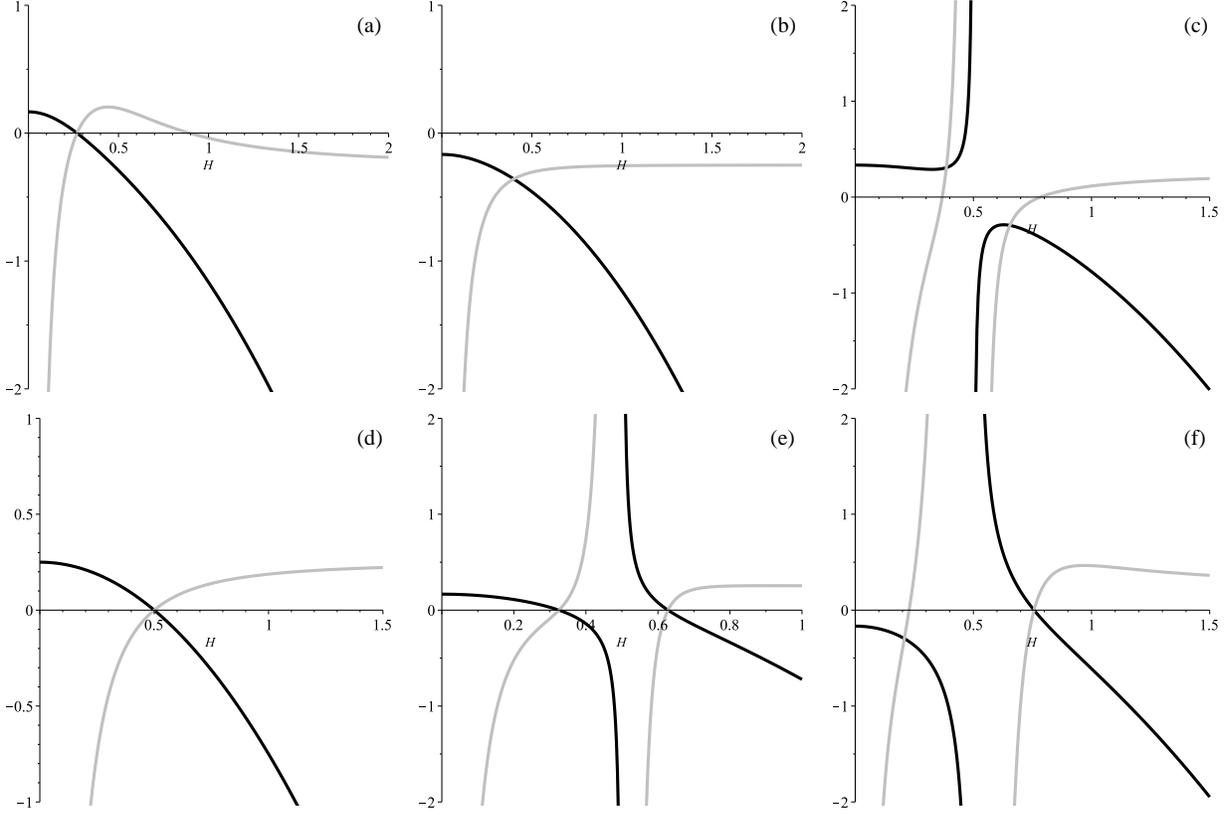}
\caption{Behavior of $\dot H(H)$ and $\dot h(H)$ from Eqs. (\ref{D1_dHdh})  in $D=1$ case. In all panels black curve corresponds to $\dot H(H)$ and grey curve to $\dot h(H)$. Panel (a) corresponds to $\alpha > 0$ and $\Lambda >0$ while panel (b) -- to $\alpha > 0$ and $\Lambda < 0$. The remaining panels correspond to $\alpha < 0$: $\alpha\Lambda < -3/2$ in (c), $\alpha\Lambda = -3/2$ in (d) and $\alpha\Lambda > -3/2$ in (e). Finally,
in (f) panel we presented typical $\alpha < 0$, $\Lambda < 0$ behavior.
(see the text for more details).}\label{D1_2}
\end{figure}

Having this in mind,
we can now plot $\dot H(H)$ and $\dot h(H)$ curves in Fig.~\ref{D1_2} and analyze them. In all panels black curves correspond to $\dot H(H)$ and grey curves to $\dot h(H)$. Panel (a) corresponds to $\alpha > 0$ and $\Lambda >0$ while panel (b) -- to $\alpha > 0$ and $\Lambda < 0$. The remaining panels correspond to $\alpha < 0$: $\alpha\Lambda < -3/2$ in (c), $\alpha\Lambda = -3/2$ in (d) and $\alpha\Lambda > -3/2$ in (e). Finally, in (f) panel we presented typical $\alpha < 0$, $\Lambda < 0$ behavior.

One important remark on the notations -- hereafter we are using notations we used in~\cite{my16a}. So we denote exponential solutions as $E$ with subindices which indicate specifics of the solution, e.g.,
$E_{iso}$ is isotropic exponential solution and so on. Power-law solutions generally denoted as $K$ as a reference to Kasner regime with $K_1$ being low-energy or GR Kasner regime, governed by $\sum p_i = 1$ and $\sum p_i^2 = 1$. High-energy or Gauss-Bonnet Kasner regime is denoted as $K_3$ as
it has $\sum p_i = 3$. Finally we denote nonstandard singularity of any nature as $nS$ -- in~\cite{my16a} we detected several different types of nonstandard singularities, and expect in current paper to have
variety of them as well, but without discrimination we denote all of them as $nS$.

Now let us have closer look on individual panels.
In (a) panel we presented behavior for $\alpha > 0$ and $\Lambda >0$ case. One can see that we have stable point which is isotropic exponential solution (since $\dot H = 0$ and $\dot h = 0$) with two
different past asymptotes. A bit further, when analyzing the behavior in Kasner exponents, we demonstrate that one of them (at $H\to\infty$) is Gauss-Bonnet Kasner $K_3$ and the other is similar to usual
(general relativity) Kasner regimes $K_1$. Since this solution is similar to $K_1$ but not exactly low-energy Kasner, we denote it as $\tilde K_1$ and discuss it later in the Discussion section.
The next (b) panel corresponds to $\alpha > 0$ and $\Lambda <0$ case. We can see that in this case $\dot H < 0$ always so it is singular transition from GB to GR Kasner-like regimes (as we demonstrate further
with Kasner exponents). All remaining panels correspond to $\alpha < 0$ case and first of them to consider is (c) panel with $\alpha\Lambda < -3/2$. There we can see two singular regimes with nonstandard
singularity at $H^2 = -1/(4\alpha)$ as future attractor: first of them, at lower $H$, has low-energy Kasner-like regime as past attractor while the second, with higher $H$, has high-energy Kasner; a bit further
we demonstrate it with Kasner exponents. Next, exact $\alpha\Lambda = -3/2$ case in (d) panel, and it is quite similar to the case in Fig.~\ref{D1_2} (a) -- the same isotropic exponential solution with two
different Kasner regimes as past attractors. The next case is more interesting and it is presented in (e) panel -- the case with $\alpha\Lambda > -3/2$. There we can see two distinct isotropic exponential
solutions -- the situation we never saw in the vacuum case~\cite{my16a}. So the the lower $H$ part we have two regimes -- $\tilde K_1$ to the first isotropic solution $E_{iso}^{(1)}$ and nonstandard singularity to
$E_{iso}^{(1)}$; the higher $H$ part is quite similar but with $K_3$ instead of $\tilde K_1$. Finally, in (f) panel we have $\alpha < 0$, $\Lambda < 0$ case. Its low-$H$ part has $nS$ to $\tilde K_1$ transition while
high-$H$ has two -- $nS$ to $E_{iso}$ and $K_3$ to $E_{iso}$.

This finalize our analysis of $\dot H(H)$ and $\dot h(H)$ curves, but to clarify Kasner regimes we perform analysis of the Kasner exponents as well. They are defined from the power-law {\it ansatz} as
$a(t) \propto t^p$ with $p$ being Kasner exponent and could be reexpressed as $p = - H^2/\dot H$. Now with use of (\ref{D1_dHdh}) we can express $p_H$ (Kasner exponent which corresponds to $H$) and $p_h$
(the same but for $h$) explicitly:

\begin{equation}
\begin{array}{l}
p_H = - \dac{H^2}{\dot H} = \dac{6H^2(1+4\alpha H^2)}{24\alpha H^4 + 12H^2 - \Lambda}, \\ \\
p_h = - \dac{h^2}{\dot h} = \dac{(1+4\alpha H^2)(\Lambda - 6H^2)^2}{576\alpha^2H^8 - 288\alpha^2\Lambda H^6 + 144\alpha H^6 - 192\alpha\Lambda H^4 + 12\alpha \Lambda^2 H^2 - 72 H^4 - 6\Lambda H^2 + \Lambda^2}.
\end{array} \label{D1_pHph}
\end{equation}

\begin{figure}
\includegraphics[width=1.0\textwidth, angle=0]{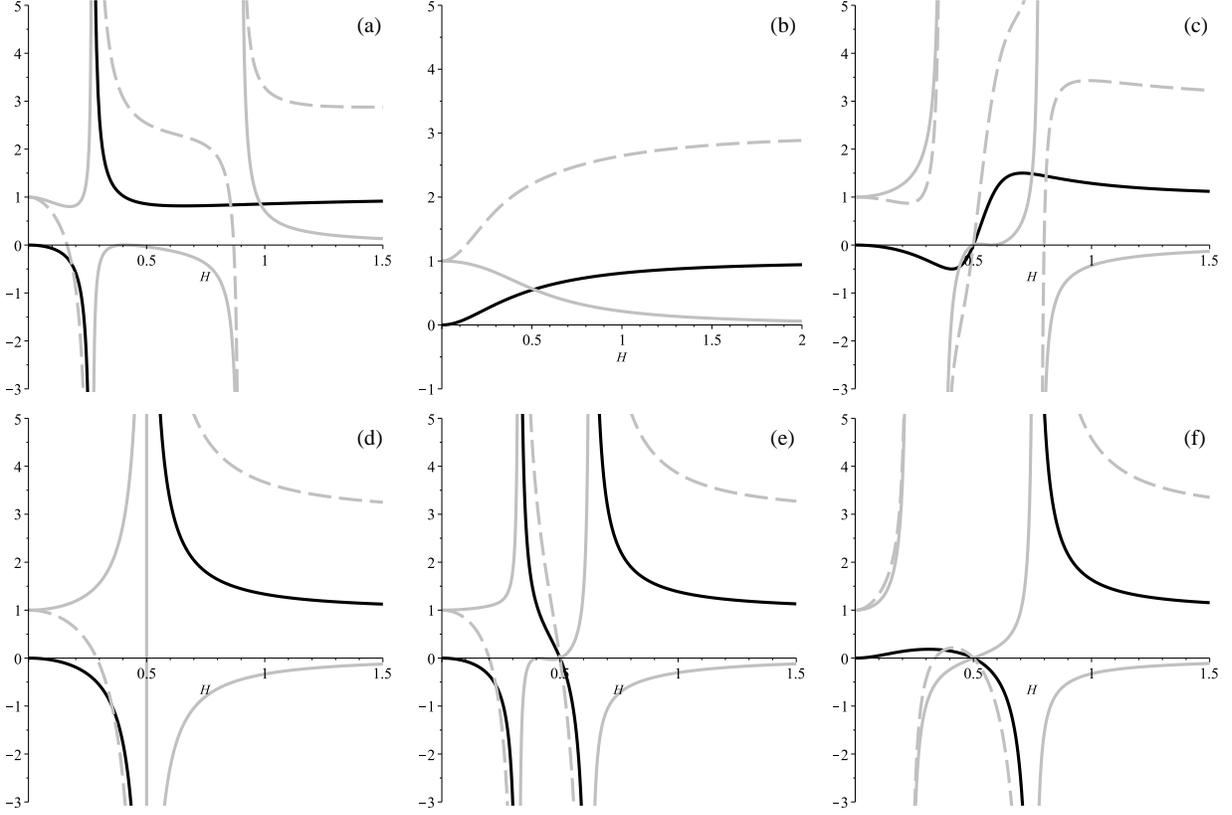}
\caption{Behavior of Kasner exponents from Eq. (\ref{D1_pHph}) in $D=1$ case. On all panels black curve corresponds to $p_H$, solid grey curve -- to $p_h$ and dashed grey -- to the expansion rate $\sum p = 3p_H + p_h$.
Panel (a) corresponds to $\alpha > 0$ and $\Lambda >0$ while panel (b) -- to $\alpha > 0$ and $\Lambda < 0$. The remaining panels correspond to $\alpha < 0$: $\alpha\Lambda < -3/2$ in (c), $\alpha\Lambda = -3/2$ in (d) and $\alpha\Lambda > -3/2$ in (e). Finally,
in (f) panel we presented typical $\alpha < 0$, $\Lambda < 0$ behavior.
(see the text for more details).}\label{D1_3}
\end{figure}

We plot the resulting curves in Fig.~\ref{D1_3}. There on all panels black curve corresponds to $p_H$, solid grey curve -- to $p_h$ and dashed grey -- to the expansion rate $\sum p = 3p_H + p_h$. The panels
layout is the same as in Fig.~\ref{D1_2} -- panel (a) corresponds to $\alpha > 0$, $\Lambda >0$; panel (b) -- to $\alpha > 0$, $\Lambda <0$; panel (c) -- to $\alpha < 0$, $\Lambda >0$ and $\alpha\Lambda < -3/2$;
panel (d) -- to $\alpha < 0$, $\Lambda >0$ and $\alpha\Lambda = -3/2$; panel (e) -- to $\alpha < 0$, $\Lambda >0$ and $\alpha\Lambda > -3/2$ and panel (f) -- to $\alpha < 0$, $\Lambda <0$.

Before comparing the dynamics in $\{\dot H, \dot h\}$ coordinates from Fig.~\ref{D1_2} with the dynamics in $\{p_H, p_h\}$ coordinates from Fig.~\ref{D1_3}, let us make several notes. First of all, from the definition of
the Kasner exponent $p = - H^2/\dot H$ any point where $\dot H = 0$ becomes singular for $p$ (except for $H=0$ sometimes), which do not corresponds to any physical singularities in $\{\dot H, \dot h\}$ description. So that the description in $\{p_H, p_h\}$ coordinates could create ``fake'' singularities. Actually, $H=0$ corresponds to the exponential solution, so divergences in $\{p_H, p_h\}$ coordinates correspond to them as well.
Secondly, again from the Kasner exponent definition, there could be a situation when at physical singularity
both $H$ and $\dot H$ diverge in a way for $p$ to remains regular -- so that the description in $\{p_H, p_h\}$ coordinates could not only create ``fake'' singularities, but also ``hide'' physical ones,
and in the vacuum case~\cite{my16a} we described several situations like that. All these makes $\{p_H, p_h\}$ description flawed, but we still use it with appropriate care to find power-law asymptotes.

The limiting values for $p_H$ and $p_h$ could be both seen from the Fig.~\ref{D1_3} and by taking appropriate limit of (\ref{D1_pHph}); one can demonstrate that for $K_1$ we have $p_H = 0, p_h = 1, \sum p = 1$
while for $K_3$ we have $p_H = 1, p_h = 0, \sum p = 3$. One can easily confirm that with Fig.~\ref{D1_3}.

As we just mentioned, exponential solutions in $\{p_H, p_h\}$ coordinates are singularities and so correspond to vertical asymptotes. Apart from them we also have zeros of $p_i$ and they correspond to
nonstandard singularities. Indeed, as we define nonstandard singularity as the situation when $\dot H$ diverges while $H$ is regular and (generally) nonzero, from the definition of the Kasner
exponent $p = - H^2/\dot H$ we can clearly see that for $\dot H = \pm\infty$ and $H\ne 0$ we have $p=0$.

So comparing Fig.~\ref{D1_2}(a) with Fig.~\ref{D1_3}(a) we can clearly see both Kasner asymptotes and detect correct values for the expansion rate for both of them ($\sum p = 1$ for $K_1$ and $\sum p = 3$
for $K_3$). Also we can see exponential solution where Kasner exponents diverge (see, e.g.,~\cite{PT} for the relations between power-law and exponential solutions) as well as ``fake'' singularity for $p_h$
where just $\dot h = 0$. Comparison of Fig.~\ref{D1_2}(b) and Fig.~\ref{D1_3}(b) clearly demonstrate $K_3 \to \tilde K_1$ transition. From the comparison of (c) panels we again can see proper power-law behavior
at $H\to 0$ and $H\to\infty$ as well as nonstandard singularity where $p_H = p_h = 0$. The (d) panels are similar to (a) -- same $\tilde K_1$ and $K_3$ separated by the exponential solution. The panel (e) of
Fig.~\ref{D1_3} is a bit more complicated -- as its counterpart from Fig.~\ref{D1_2} -- we detect $\tilde K_1$ and $K_3$, two exponential solutions and nonstandard singularity ``between'' them -- all from both
Fig.~\ref{D1_2}(e) and Fig.~\ref{D1_3}(e). Finally, in (f) panel, apart from $\tilde K_1$ and $K_3$ we can see exponential solution and nonstandard singularity. All this comparison demonstrates that the descriptions
in $\{\dot H, \dot h\}$ and $\{p_H, p_h\}$ coordinates correspond to each other and adequately describe the dynamics of the system.

\begin{table}[h]
\begin{center}
\caption{Summary of $D=1$ $\Lambda$-term regimes.}
\label{D.1}
  \begin{tabular}{|c|c|c|c|c|}
    \hline
     $\alpha$ & $\Lambda$ & \multicolumn{2}{c|}{Additional conditions} & Regimes  \\
    \hline
\multirow{3}{*}{$\alpha > 0$} & \multirow{2}{*}{$\Lambda > 0$} & \multicolumn{2}{c|}{$H < H_-$ from (\ref{D1_dHdh_0})} & $\tilde K_1 \to E_{iso}$ \\  \cline{3-5}
& & \multicolumn{2}{c|}{$H > H_-$ from (\ref{D1_dHdh_0})} & $K_3 \to E_{iso}$ \\  \cline{2-5}
& $\Lambda < 0$ & \multicolumn{2}{c|}{no} & $K_3 \to \tilde K_1^S$ \\  \cline{1-5}
\multirow{11}{*}{$\alpha > 0$} & \multirow{8}{*}{$\Lambda > 0$} & \multirow{2}{*}{$\alpha\Lambda < -3/2$} & $H < \frac{1}{2\sqrt{-\alpha}}$ & $\tilde K_1 \to nS$ \\ \cline{4-5}
& & & $H > \frac{1}{2\sqrt{-\alpha}}$ & $K_3 \to nS$ \\ \cline{3-5}
& & \multirow{2}{*}{$\alpha\Lambda = -3/2$} & $H < \frac{1}{2\sqrt{-\alpha}}$ & $\tilde K_1 \to E_{iso}$ \\ \cline{4-5}
& & & $H > \frac{1}{2\sqrt{-\alpha}}$ & $K_3 \to E_{iso}$ \\ \cline{3-5}
& & \multirow{4}{*}{$\alpha\Lambda > -3/2$} & $H < H_-$ from (\ref{D1_dHdh_0}) & $\tilde K_1 \to E_{iso}^{(1)}$ \\ \cline{4-5}
& & & $\frac{1}{2\sqrt{-\alpha}} > H > H_-$ from (\ref{D1_dHdh_0}) & $nS \to E_{iso}^{(1)}$ \\ \cline{4-5}
& & & $H_+ > H > \frac{1}{2\sqrt{-\alpha}}$ from (\ref{D1_dHdh_0}) & $nS \to E_{iso}^{(2)}$ \\ \cline{4-5}
& & & $H > H_+$ from (\ref{D1_dHdh_0}) & $K_3 \to E_{iso}^{(2)}$ \\ \cline{2-5}
& \multirow{3}{*}{$\Lambda < 0$} & \multicolumn{2}{c|}{$H < \frac{1}{2\sqrt{-\alpha}}$} & $nS \to \tilde K_1^S$ \\ \cline{3-5}
& & \multicolumn{2}{c|}{$H_+ > H > \frac{1}{2\sqrt{-\alpha}}$ from (\ref{D1_dHdh_0})} & $nS \to E_{iso}$ \\ \cline{3-5}
& & \multicolumn{2}{c|}{$H > H_+$ from (\ref{D1_dHdh_0})} & $K_3 \to E_{iso}$ \\
    \hline
  \end{tabular}
\end{center}
\end{table}

Now let us summarize all regimes found in Table~\ref{D.1}. We can see a variety of different regimes -- much more than we have in $D=1$ vacuum regime~\cite{my16a}. Almost all of the regimes are singular
with the only nonsingular regime $K_3 \to E_{iso}$. But we cannot call it viable -- indeed, isotropisation here means equality of all four space dimensions which clearly violate our observations. So
we report that, despite of a variety of regimes presented in Table~\ref{D.1}, none of them is viable -- the situation we never had in vacuum case. Also we need to stress readers' attention on power-law
solutions once again -- as we noted, $\tilde K_1$ is a regime similar to GR Kasner regime $K_1$ but singular -- indeed, from Fig. \ref{D1_1} one can see that upon approaching $H=0$ we have divergence in
$\dot h$ - so that $H=0$ is singular point while ``true'' Kasner regime (in Bianchi-I, where it was originally defined) is nonsingular. But as $H=0$ is approached in ``nearly Kasner'' manner, we denote
this regime as $\tilde K_1$. When this regime is the past attractor, we denote it as $\tilde K_1^S$ to stress that it is singular, unlike ``usual'' $K_1$ Kasner we saw in the vacuum case~\cite{my16a}.
All these issues are discussed in Discussion section.

\section{$D=2$ case}

In this case the equations of motion take the form ($H$-equation, $h$-equation, and constraint correspondingly)

\begin{equation}
\begin{array}{l}
4\dot H + 6H^2 + 4\dot h + 6h^2 + 8Hh + 8\alpha \( 2(\dot H + H^2) (2Hh + h^2) + 2(\dot h + h^2) (H^2 + 2Hh) +3H^2h^2\)  = \Lambda,
\end{array} \label{D2_H}
\end{equation}

\begin{equation}
\begin{array}{l}
6\dot H + 12H^2 + 2\dot h + 2h^2 + 6Hh + 8\alpha \( 3(\dot H + H^2) (H^2 + 2Hh) + 3(\dot h + h^2)H^2 + 3H^3h  \)  = \Lambda,
\end{array} \label{D2_h}
\end{equation}

\begin{equation}
\begin{array}{l}
6H^2 + 12Hh + 2h^2 + 24\alpha (2H^3h + 3H^2h^2 ) = \Lambda.
\end{array} \label{D2_con}
\end{equation}

If we solve (\ref{D2_con}) with respect to $h$ we get

\begin{equation}
\begin{array}{l}
h_\pm = - \dac{ 24\alpha H^3 + 6H \pm \sqrt{ 576\alpha^2 H^6 - 144\alpha H^4 + 24H^2(1+3\alpha\Lambda) + 2\Lambda } }{2(1+36\alpha H^2)}.
\end{array} \label{D2_hh}
\end{equation}

\begin{figure}
\includegraphics[width=1.0\textwidth, angle=0]{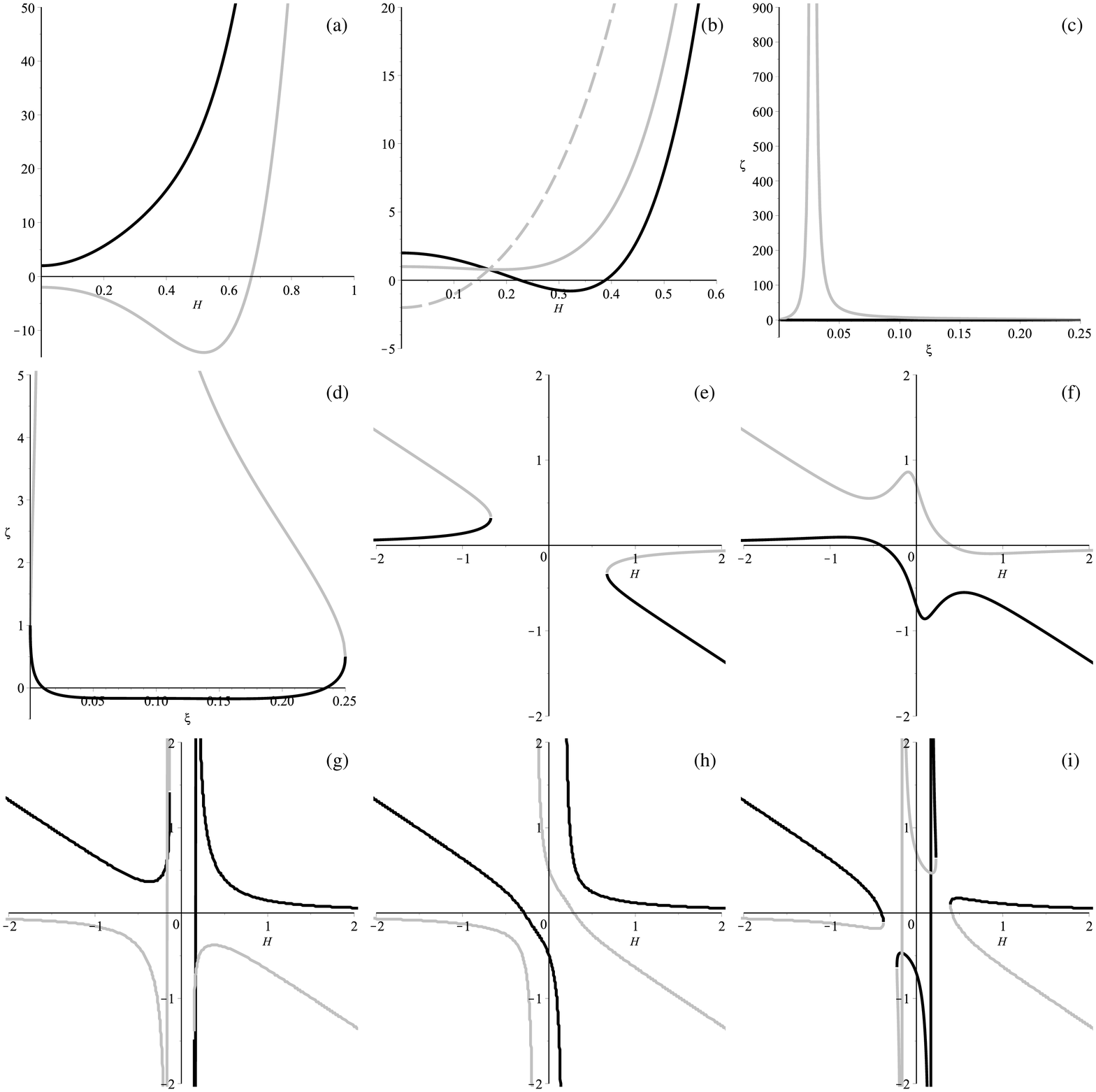}
\caption{Behavior of the radicand from (\ref{D2_hh}) in (a) and (b) panels, reduced denominator (\ref{D2_zn}) in (c) and (d) panels and $h(H)$ curves from (\ref{D2_hh}) for different cases in (e)--(i) panels for $D=2$ case.
(see the text for more details).}\label{D2_0}
\end{figure}

Let us first have a closer look on the radicand in (\ref{D2_hh}). It is bicubic equation with discriminant

\begin{equation}
\begin{array}{l}
\Delta = - 3981312(6\xi + 1)^2(6\xi + 5),
\end{array} \label{D2_hh_discr}
\end{equation}

\noindent where $\xi = \alpha\Lambda$. As we know, if the discriminant of the cubic equation is positive -- it has three real roots, if negative -- only one real root. So that for $\xi > -5/6$ we have one and for
$\xi < -5/6$ we have three roots. But since the equation is bicubic, not only the number but the sign of the roots is important, so we plot in Fig.~\ref{D2_0}(a, b) behavior of the radicand from (\ref{D2_hh}) for $H>0$. The cases presented on panel (a) are for $\alpha > 0$: $\Lambda > 0$ as black line and $\Lambda < 0$ as grey line. One can see that the former of them has $H>0$ as a domain of definition (which defined from the positivity of the radicand) while the latter
has only $H > H_1 > 0$. On the next, (b) panel, we presented radicand behavior for
$\alpha < 0$ case: $\Lambda > 0$, $\alpha\Lambda < -5/6$ as black line, $\Lambda > 0$, $\alpha\Lambda > -5/6$ as solid grey line and $\Lambda < 0$ as dashed grey line. One can see that, similar to $D=1$ case,
only $\alpha < 0$ cases are affected by $\alpha\Lambda = -5/6$ separation. So we can see that the $\Lambda > 0$, $\alpha\Lambda < -5/6$ case has twofold discontinuous domain of definition -- $0<H<H_1$ and $H>H_2$
with both $H_1$ and $H_2$ defined from the roots of the radicand. The $\Lambda > 0$, $\alpha\Lambda > -5/6$ case has entire $H>0$ as a domain of definition and $\Lambda < 0$ case has $H > H_1 > 0$, similar to
$\alpha > 0$, $\Lambda < 0$ case.

With that at hand we can plot $h(H)$ curves from (\ref{D2_hh}). They are presented in (e)--(i) panels of Fig.~\ref{D2_0}. On all panels black curve corresponds to $h_+$ while grey -- to $h_-$. So on (e) panel
we presented $\alpha > 0$, $\Lambda < 0$ case -- and according to (a) panel, its domain of definition is $H > H_1 > 0$. One can also see that branches ``turn'' into each other, making impossible entire
evolution -- we demonstrate it more explicitly later. On (f) panel we presented $\alpha > 0$, $\Lambda > 0$ case and it has entire $H>0$ as its domain of definition; branches are also separated from each other.
Remaining three panels correspond to $\alpha < 0$: $\Lambda < 0$ on (g), $\Lambda > 0$, $\alpha\Lambda > -5/6$ on (h) and $\Lambda > 0$, $\alpha\Lambda < -5/6$ on (i). One can clearly see that their domains of
definition are in accordance with radicand classification. Let us make a note similar to the $D=1$ case -- for successful compactification one needs $H>0$ and $h\leqslant 0$ so that from Fig.~\ref{D2_0} one can judge about the
regions of the initial conditions and parameters where it could be satisfied.

Now with $h(H)$ behavior described, we can turn our attention to $\dot H(H)$ and $\dot h(H)$ behavior. We solve (\ref{D2_H})--(\ref{D2_h}) with respect to $\dot H$ and $\dot h$ and substitute (\ref{D2_hh}) to
get the following expressions:

\begin{equation}
\begin{array}{l}
\dot H_\pm = - \dac{H^2 P^H_\pm}{4Q_\pm}, \\
\dot h_\pm = \dac{3H^2 P^h_\pm}{4(1+36\alpha H^2)^2 Q_\pm}
\end{array} \label{D2_dHdh_1}
\end{equation}

\noindent with

\begin{equation}
\begin{array}{l}
P^H_\pm = 96768\xi^5 + (\mp 1152 \mathcal{D} - 19584)\xi^4 + (\mp 1248\mathcal{D} + 12096\zeta + 2592)\xi^3 + \\ + (\pm 144\zeta \mathcal{D} \mp 64\mathcal{D} + 912\zeta + 264)\xi^2 + (\pm 4\zeta \mathcal{D}
\mp 4 \mathcal{D} - 20\zeta - 12)\xi - \zeta; \\
P^h_\pm = 9953280\xi^7 + (-9787392\pm 414720 \mathcal{D})\xi^6 + (580608 + 2737152\zeta \mp 76032\mathcal{D})\xi^5 + \\ + (\mp 51840\zeta \mathcal{D} \mp 12096\mathcal{D} + 449280\zeta + 158976)\xi^4 +
(\mp8352\zeta\mathcal{D} - 62208\zeta^2 \mp 1296 \mathcal{D} + \\ + 3456\zeta + 672)\xi^3 + (\pm168\zeta \mathcal{D} - 3456\zeta^2 \pm 160\mathcal{D} + 3552\zeta + 1064)\xi^2 + \\ + (\pm 10\zeta \mathcal{D}
- 48\zeta^2 \mp 6\mathcal{D} + 68\zeta - 28)\xi - \zeta; \\
Q_\pm = 31104\xi^4 - 2880\xi^3 + (1296\zeta + 216 \mp 192\mathcal{D})\xi^2 + (\mp 16\mathcal{D} - 12)\xi - \zeta + 1;
\end{array} \label{D2_dHdh_2}
\end{equation}

\noindent where $\xi = \alpha H^2$, $\zeta = \alpha\Lambda$ and $\mathcal{D} = \sqrt{576\xi^2 + 72\zeta - 144\xi + 24 + 2\zeta/\xi}$.
Before moving forward it is useful to analyze (\ref{D2_dHdh_2}) and find their roots to locate zeros and asymptotes of $\dot H$ and $\dot h$. One can demonstrate that $P^H_+$ could be rewritten in a form

\begin{equation}
\begin{array}{l}
P^H_+ \Leftrightarrow (120\xi^2 + 20\xi - \zeta)(192\xi^3 - 112\xi^2 + 4\xi(1+8\zeta) - 1)(288\xi^3 - 72\xi^2 + 12\xi(1+3\zeta) + \zeta),
\end{array} \label{D2_ch_dH}
\end{equation}

\noindent so it have up to six real roots in five regions divided by four isolated points: \mbox{$\xi = \{-5/6, -1/6, 15/32, 1/2\}$}. Its counterpart $P^H_-$ has the same roots plus an additional root at $\xi=-1/36$.
Similarly, one can demonstrate that $P^h_+$ could be rewritten in a form

\begin{equation}
\begin{array}{l}
P^h_+ \Leftrightarrow (120\xi^2 - 20\xi - \zeta)(120\xi^2 + 20\xi - \zeta)(144\xi^2 - 36\xi(1+2\zeta) + 1)\times \\ \times (192\xi^3 - 112\xi^2 + 4\xi(1+8\zeta) - 1),
\end{array} \label{D2_ch_dh}
\end{equation}

\noindent and it have up to nine real roots in five regions divided by the same four isolated points: \mbox{$\xi = \{-5/6, -1/6, 15/32, 1/2\}$}. Similarly to $P^H_-$, $P^h_-$ has the same as $P^h_+$ roots
plus an additional root at $\xi=-1/36$.

As joint roots $\dot h=0$ and $\dot H=0$ determine the locations of exponential solutions, it is important to find them. Comparison of (\ref{D2_ch_dH}) with (\ref{D2_ch_dh}) clearly pinpoints the equation that
governs the locations of the exponential solutions:

\begin{equation}
\begin{array}{c}
(120\xi^2 + 20\xi - \zeta) \times (192\xi^3 - 112\xi^2 + 4\xi(1+8\zeta) - 1) = 0 \\
\Updownarrow \\
\xi_\pm = - \dac{1}{12} \pm \dac{\sqrt{30\zeta + 25}}{60}, \\ \\ \xi_1 = - \dac{1}{36} \( -756\zeta + 370 + 18\sqrt{ 1152\zeta^3 - 156\zeta^2 - 660\zeta + 225 } \)^{1/3} - \\ \\ -
\dac{18\zeta - 10}{9 \( -756\zeta + 370 + 18\sqrt{ 1152\zeta^3 - 156\zeta^2 - 660\zeta + 225 } \)^{1/3}} + \dac{7}{36},~\xi_{2,\,3},
\end{array} \label{D2_exp}
\end{equation}

\noindent where $\xi_\pm$ are roots of quadratic equation, $\xi_1$ is the first (always real) root of the cubic equation while $\xi_2 > \xi_3$ are two remaining roots of cubic equation. Analyzing them
separately for positive and negative $\alpha$ and $\Lambda$, we can conclude: for $\alpha > 0$, $\Lambda > 0$ we have $\xi_+$ governs location of the exponential solution for $h_-$ while roots of cubic
equation govern locations of exponential solutions for $h_+$: there is one solution for $\zeta < 15/32$ and $\zeta > 1/2$, two solutions for $\zeta = 15/32,\, 1/2$ and three solutions for $15/32 < \zeta < 1/2$.
For $\alpha > 0$ and $\Lambda < 0$ we have one exponential solution for $h_+$ which is governed by $\xi_-$. For $\alpha < 0$, $\Lambda > 0$ the situation is following: for $\zeta > -5/6$ we have two exponential solutions governed by
$\xi_\pm$; for $\zeta = -5/6$ there remains only one solution (at that point these two roots coincide) and for $\zeta < -5/6$ there are two exponential solutions governed by $\xi_{2,\,3}$ (for $\zeta \to -5/6+0$
$\lim \xi_{2,\,3} = \lim \xi_\pm$ so the roots smoothly transit into each other). One of these roots corresponds to $h_-$ and the other -- to $h_+$.
Finally, for $\alpha < 0$ and $\Lambda < 0$ we have only one exponential solution governed by $\xi_1$ and it appears in $h_+$ branch.

Finally, let us consider the denominator $Q_\pm$. It could be reduced to sixth order algebraic equation with respect to $\xi$ but unlike numerators cannot be solved explicitly. But we could solve $Q_+$ with respect to $\zeta$:

\begin{equation}
\begin{array}{l}
\zeta_\pm = - \dac{31104\xi^4 - 4608\xi^3 - 600\xi^2 - 172\xi - 1 \pm 16(12\xi + 1)^2\sqrt{-2\xi(4\xi-1)}}{1296\xi^2 - 72\xi + 1}.
\end{array} \label{D2_zn}
\end{equation}

We plot both branches -- $\zeta_+$ as black and $\zeta_-$ as grey -- in Fig.~\ref{D2_0}(c, d). There on (c) panel we presented large-scale structure of $\zeta_-$ and in (d) -- fine structure of $\zeta_+$. One can
see that $\zeta_-$ never crosses zero while $\zeta_+$ do. Also, similar to numerators, $Q_-$ has additional root at $\xi=-1/36$. Further analyzing $Q_\pm$ and (\ref{D2_zn}) leads us to the following conclusions:
for $\alpha > 0$ and $\Lambda < 0$, $Q_+$ does not have roots for $\zeta < \zeta_0 \approx -0.17359961$\footnote{This value was obtained numerically -- as the original equation is sixth order it is impossible to solve it in radicals.}, have one for $\zeta = \zeta_0$ and have two for $\zeta > \zeta_0$; $Q_-$ has no roots for $\zeta < 1$ and $\alpha > 0$, has one at $H=0$ at $\zeta=1$ and one at $H > 0$ at $\zeta > 1$. At $\alpha < 0$ neither of $Q_\pm$ have zeros. For future use we denote $\xi_4 > \xi_5$ as Eq. (\ref{D2_zn}) roots: to each particular $\zeta$ they could be found numerically.

\begin{figure}
\includegraphics[width=1.0\textwidth, angle=0]{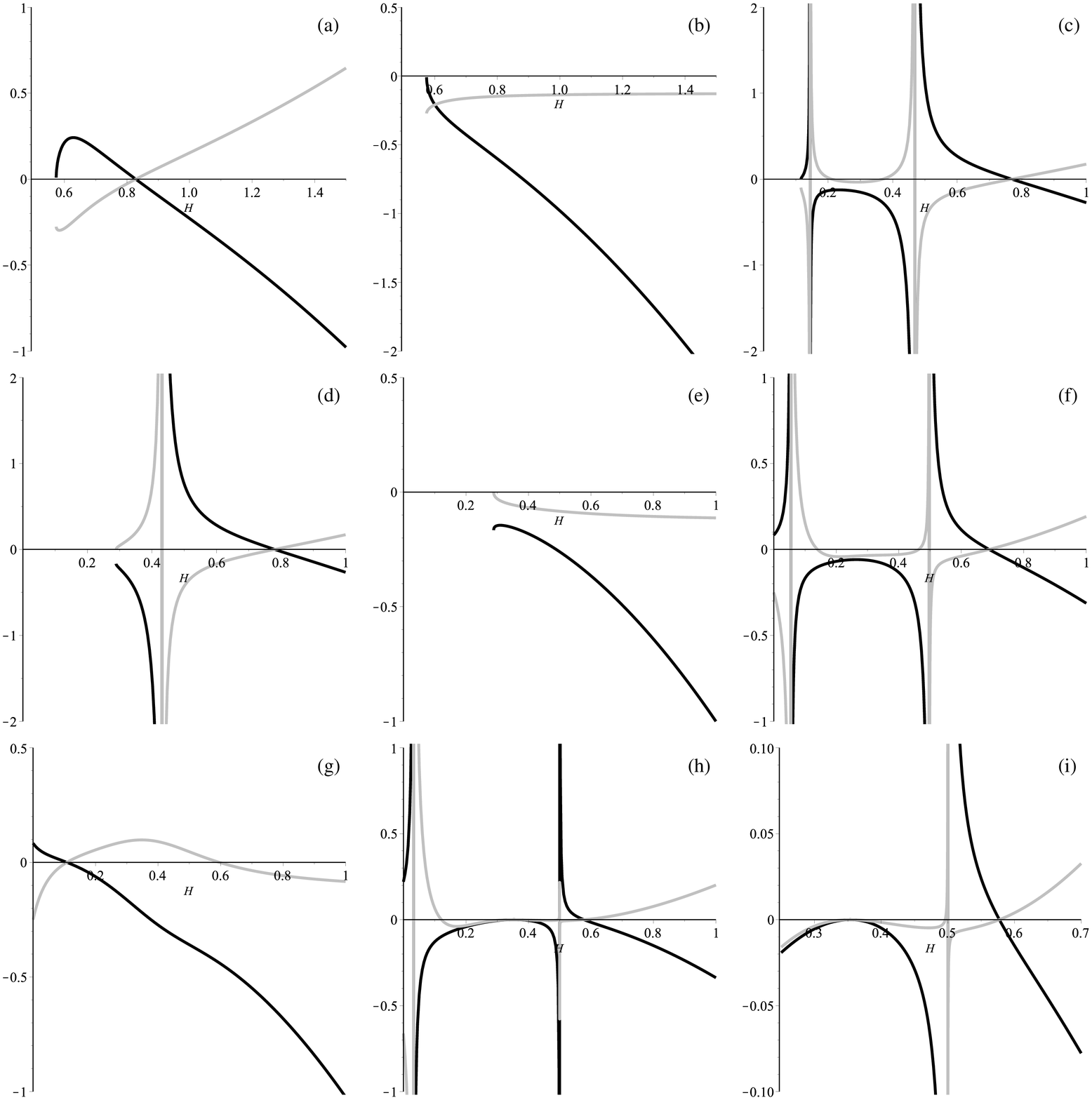}
\caption{Behavior of $\dot H(H)$ (black curve) and $\dot h(H)$ (grey curve) in $D=2$ case for $\alpha > 0$ and: $\zeta < \zeta_0$ on (a) and (b) panels; $0 > \zeta > \zeta_0$, $\zeta\ne -1/6$ on (c) panel; $\zeta = -1/6$ on (d) and (e) panels;
$0 < \zeta < 15/32$ on (f) and (g) panels; $\zeta = 15/32$ on (h) and (i) panels
(see the text for more details).}\label{D2_1}
\end{figure}

With all this preliminary considerations done, it is time to present $\dot H(H)$ and $\dot h(H)$ curves for all possible distinct areas of parameters. We presented these curves in Figs.~\ref{D2_1}--\ref{D2_3}.
In all these figures black curves correspond to $\dot H(H)$ while grey -- to $\dot h(H)$. Let us have a closer look on all panels in all figures.

In Fig.~\ref{D2_1}(a, b) we presented the situation for $\alpha > 0$ and $\zeta < \zeta_0$ mentioned above \mbox{($\zeta_0 \approx -0.17359961$)} -- $h_+$ in (a) panel and $h_-$ in (b). We can see that the 
domain of definition does not reach $H=0$, all
according to $h(H)$ graphs  investigation. We can see anisotropic exponential solution in (a) panel -- it could be verified by taking appropriate root from (\ref{D2_ch_dH}) and (\ref{D2_ch_dh}) and 
substituting it into $h_+$. Apart from it, we can see $K_3$ regime at high $H$ and nonstandard singularity. On (b) panel we can see the singular transition from $K_3$ to another $nS$. On panel (c) we 
presented behavior for $h_+$ for $\alpha > 0$ and $0 > \zeta > \zeta_0$, $\zeta\ne 1/6$. We can see more complicated behavior there -- some singular behavior between two different singularities, again 
singular behavior between two different nonstandard singularities followed by $nS$ to anisotropic exponential solution transition and $K_3$ to this exponential solution. Another branch $h_-$
has the same behavior as in $\zeta < \zeta_0$ case so we do not present it again. On the following (d) and (e) panels we present $\zeta= -1/6$ case -- $h_+$ branch on (d) and $h_-$ branch on (e). On the 
former of them we
can see singular transition followed by $nS$ to anisotropic exponential regime and $K_3$ to anisotropic exponential solution transitions. On (e) panel we see transition from $K_3$ to $nS$. This finalize our study of
$\alpha > 0$ $\Lambda < 0$ regimes and we turn to $\alpha > 0$ $\Lambda > 0$ ones. First of them presented in Fig.~\ref{D2_1}(f, g) and it is $0 < \zeta < 15/32$ case: on (f) panel we have $h_+$ and 
on (g) $h_-$ branches. We see that
turning to $\Lambda > 0$ changed the domain of definition and now it is entire $H>0$. So the regimes on (f) panels are: some power-law to nonstandard singularity, $nS$ to $nS$ and anisotropic exponential solution
as future attractor for $nS$ and $K_3$. But what we see as some power-law regime at low $H$ in reality is something else -- indeed, from  Fig. \ref{D2_0}(e)--(i) one can see that in this case (and some
other further cases) $H=0$ is regular and nonzero point for $h$. So that $H=0$ is not an endpoint and the evolution must be prolonged to $H<0$ domain. As (\ref{D2_dHdh_1})--(\ref{D2_ch_dh}) are symmetric
with respect to $H=0$ (they contain only even powers of $H$), one can restore entire $\dot H(H)$ and $\dot h(H)$ graphs by mirroring $H>0$ part with respect to $H=0$. Then we can see that the past attractor
in this case is nonstandard singularity $nS^{(-)}$, dual to the closest to $H=0$ nonstandard singularity.
On (g) panel we have the behavior for $h_-$ which is typical for all $1 > \zeta > 0$ -- isotropic exponential solution as future attractor
and $K_3$ for large $H$. For small $H$, with the same reasoning as in (f) case, we retrieve $E_{iso}^{(-)}$ -- isotropic shrinking, dual to exponential expansion $E_{iso}$ from future attractor. So that
in a sense this regime is a bounce from isotropic exponential contraction to isotropic exponential expansion.
Finally on (h) and (i) panels of Fig.~\ref{D2_1} we presented behavior for $h_+$ for $\zeta = 15/32$ (as we mentioned, $h_-$ branch has behavior similar to presented on (g) panel). From (h) panel
one can see that it resemble (f) panel, but the top between nonstandard singularities now ``touches'' $\dot H = 0$ line, as seen on (i) panel, creating second anisotropic solution between two nonstandard singularities. So now the
solutions include $nS^{(-)}$ to $nS$ at lowest $H$, then exponential solution to $nS$ transition, $nS$ to this exponential solution transition (this exponential solution at $H=H_0$ has
directional stability -- it is stable for $H \to H_0 + 0$ but unstable for $H \to H_0 - 0$), $nS$ to another anisotropic exponential solution and $K_3$ to this solution. So in this case we have two
different anisotropic solutions -- one of them has directional stability while another is stable.

\begin{figure}
\includegraphics[width=1.0\textwidth, angle=0]{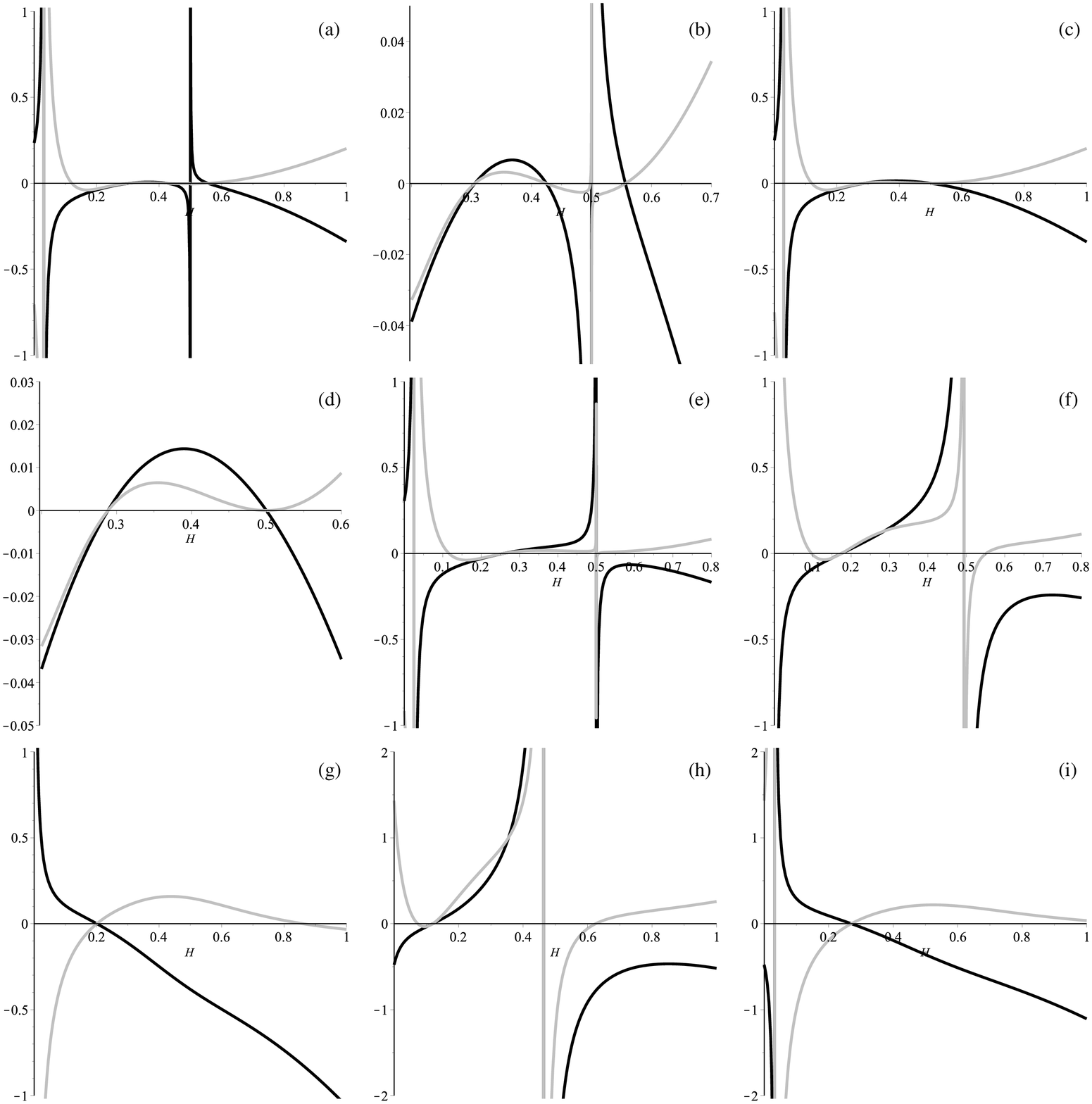}
\caption{Continuation of Fig.~\ref{D2_1} for $\alpha > 0$ and: $1/2 > \zeta > 15/32$ on (a) and (b) panels; $\zeta = 1/2$ on (c) and (d) panels; $1 > \zeta > 1/2$ on (e) panel; $\zeta = 1$ on (f) and (g) panels;
$\zeta > 1$ on (h) and (i) panels
(see the text for more details).}\label{D2_2}
\end{figure}

The description of all regimes continues in Fig.~\ref{D2_2}. There on (a) and (b) panels we presented the behavior for $1/2 > \zeta > 15/32$. Comparing them with Fig.~\ref{D2_1}(f, g) one can see that the ``touch''
from $\zeta = 15/32$ case now moved upper creating two exponential solutions, as seen on (b) panel. So in this case the regimes are: $nS^{(-)}$ to $nS$ at lowest $H$, the first exponential solution to $nS$,
the first exponential solution to the second exponential solution, $nS$ to the second exponential solution, $nS$ to the third exponential solution and $K_3$ to it. We see that in that case we have three different
anisotropic exponential solutions, and all of them have $H>0$ with $h<0$. This is so far the most exceptional situation which we never saw in the vacuum case~\cite{my16a}. Of these three exponential
solutions, the first one is unstable while the second and third are stable. Our description continues to the $\zeta = 1/2$ case presented on panels (c) and (d). We can see
that the second nonstandard singularity is ``gone'' (or, rather, ``moved'' to $H=\infty$) so we have $nS^{(-)}$ to $nS$ at lowest $H$, first exponential solution to $nS$, first exponential solution to the
second exponential solution and $K_3$ to the second exponential solution. Next is the $1 > \zeta > 1/2$ case which is presented in Fig.~\ref{D2_2}(e). We see that the second nonstandard singularity is back
and the regimes include
$nS^{(-)}$ to $nS$ at lowest $H$, exponential solution to $nS$, the same exponential solution to another $nS$ and $K_3$ to that $nS$ -- so all the regimes are singular in that case. Next case to consider
is $\zeta = 1$, $\dot H(H)$ and $\dot h(H)$ for it are presented on (f) and (g) panels. As we mentioned, in this case we have nonstandard singularity at $H=0$ so the regimes for $h_+$ branch, presented on (f) panel,
are exponential to $nS$, exponential to another $nS$ and $K_3$ to $nS$ - same as above, we have no regimes with nonsingular future asymptote. On (g) panel we presented the behavior for $h_-$; one can see that it
is quite similar to Fig.~\ref{D2_1}(g) with the difference that at $H=0$ we now have nonstandard singularity instead of continuation to $H<0$. Finally, on (h) and (i) panels we present $\zeta > 1$ regimes. As we
mentioned while describing the denominator of $\dot H$ and $\dot h$, now $h_-$ branch could have zero in denominator and so nonstandard singularity appears on (i) panel.
In some sense, one of two nonstandard singularities (see e.g. Fig.~\ref{D2_2}(e)) ``moved'' from $h_+$ branch to $h_-$ and now both $h_+$ ((h) panel) and $h_-$ ((i) panel) branches have $nS$.
So for $h_+$ branch the regimes are expanding exponential to contracting exponential (derived with the same argumentation as $nS^{(-)}$ in previous cases), exponential to $nS$ and $K_3$ to $nS$.
For $h_-$ branch, the regimes are $nS$ to $nS^{(-)}$, $nS$ to isotropic exponential solution and $K_3$ to this solution. This finalize our study of $\alpha > 0$ regimes, on the remaining Fig.~\ref{D2_3}
we collected $\alpha < 0$ cases.

\begin{figure}
\includegraphics[width=1.0\textwidth, angle=0]{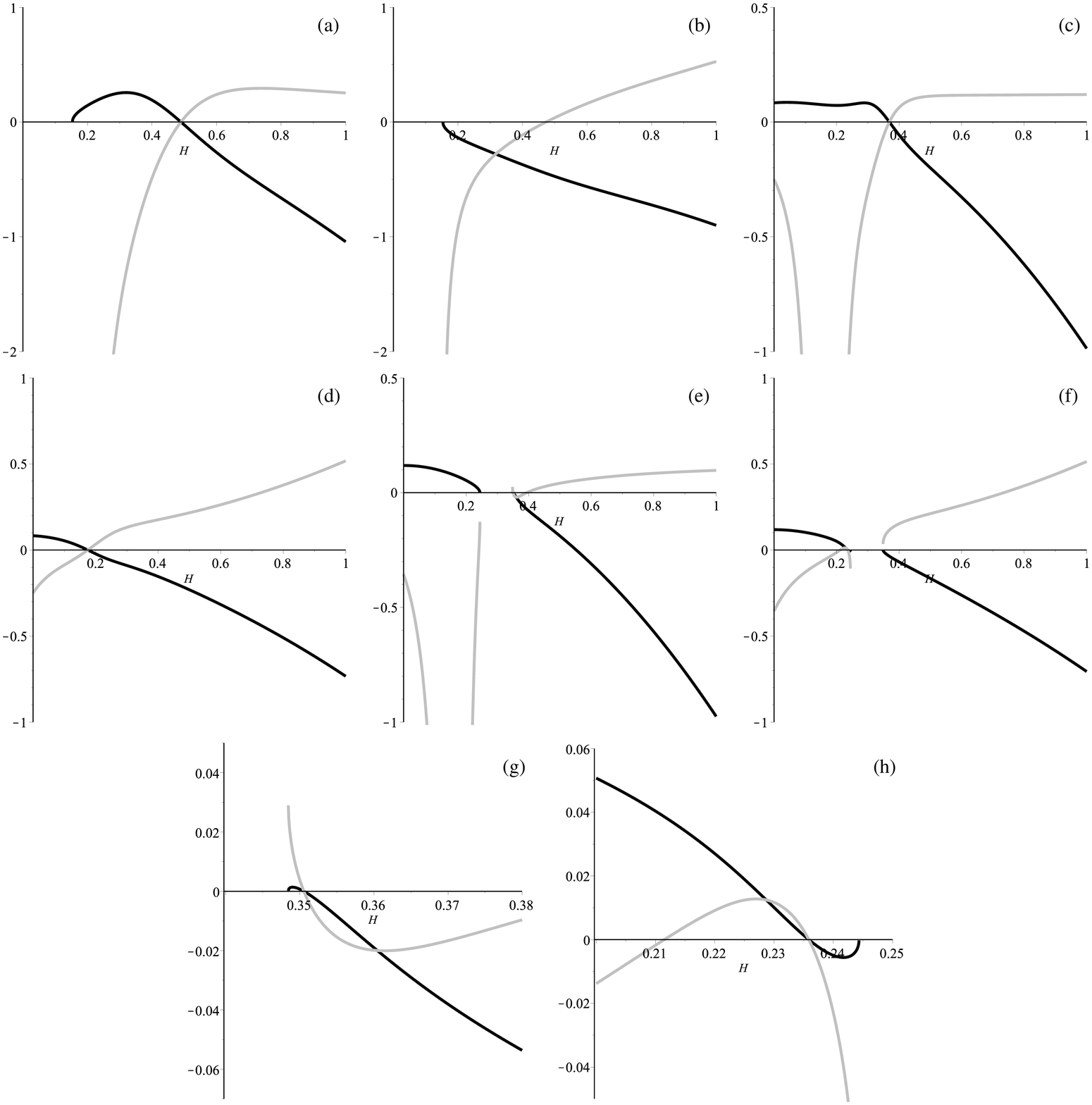}
\caption{Continuation of Figs.~\ref{D2_1} and \ref{D2_2} for $\alpha < 0$ and: $\zeta > 0$ on (a) and (b) panels; $0 > \zeta > -5/6$ on (c) and (d) panels; $\zeta < -5/6$ on (e)--(h) panels
(see the text for more details).}\label{D2_3}
\end{figure}

So the first of $\alpha < 0$ cases to consider is the case $\zeta > 0$ presented in Fig.~\ref{D2_3}(a, b). We can see that, according to the $h(H)$ description, its domain of definition does not reach $H=0$ and
the regimes resemble those of $\alpha > 0$ case (compare with Fig.~\ref{D2_1}(a, b)). So the regimes for $h_+$ branch, presented on (a) panel, are $nS$ to isotropic exponential solution and $K_3$ to the
same exponential solution. Another branch, $h_-$, presented on (b) panel, demonstrates only $K_3$ to $nS$ transition. The next distinct behavior corresponds to $0 > \zeta > -5/6$ and it presented in (c) and (d) panels -- $h_+$ on
(c) and $h_-$ on (d). We can see that the domain of definition covers entire $H>0$. For $h_+$ the regimes are $nS^{(-)}$ to $nS$ (with the same reasoning as in previous cases), $nS$ to isotropic exponential and $K_3$ to the same regime while for $h_-$
they are isotropic exponential contraction to isotropic exponential expansion (similar to presented in
 Fig.~\ref{D2_1}(g)), and $K_3$ to the isotropic exponential expansion. Finally, last four panels correspond to $\zeta < -5/6$ case. From (e) and (f) panels we can see that it is the case with two-fold
discontinuous domain of definition -- we have one domain at lower $H$ and another at higher. Panel (e) corresponds to $h_+$ while (f) -- to $h_-$. Also, panel (g) shows fine structure of the feature on the
higher-$H$ branch from panel (e) while panel (h) -- similar feature but from lower-$H$ branch from panel (f). With all these taken into account, the list of regimes for $h_+$ branch includes $nS^{(-)}$
 to $nS$ at low $H$, $nS$ to exponential solution, and $K_3$ to the same exponential solution. The regimes for $h_-$ include exponential contraction to expansion -- opposite to what we see in Fig.~\ref{D2_2}(h),
 $nS$ to the same exponential solution and $K_3$ to another $nS$. At this point it worth mentioning that all exponential solutions in this case are anisotropic, but if we take $h(H)$ expressions and figures into account (Fig.~\ref{D2_0}),
we can note that both $H>0$ and $h>0$ and in some cases we could even have $h>H$, so that it is unlikely for these solutions to give us proper dynamical compactification.

The situation changes at $\zeta \leqslant -3/2$ -- at this point the exponential solution from $h_-$ branch disappears, leaving $h_-$ branch with just nonstandard singularities as future asymptotes. As of
$h_+$ branch, at $\zeta = -3/2$ we have $h=0$ and for $\zeta < -3/2$ we have $h>0$ so the corresponding transition from the high-energy Kasner to the exponential solution becomes viable.

Finally, to complete description of this case we need to consider the dynamics in Kasner exponents. Being defined as $p = -H^2/\dot H$, they could be expressed through $h(H)$ (see Eq.(\ref{D2_hh})), 
$\dot H$ and $\dot h$ (see Eqs. (\ref{D2_dHdh_1})--(\ref{D2_dHdh_2})) with $p_H$ being Kasner exponent associated with $H$ and $p_h$ -- with $h$. The analysis of $\dot H$ and $\dot h$ roots and  asymptotes 
is applicable to $p_H$ and $p_h$ as well, so we present the resulting behavior of $p_H$ and $p_h$ in Figs.~\ref{D2_4}--\ref{D2_7}.

The analysis of these curves is quite similar to how we did it in $D=1$ case, so let us skip minor details. In Fig.~\ref{D2_4} we presented cases for $\alpha > 0$ and $\zeta < \zeta_0$ on (a) and (b) 
panels; $0 > \zeta > \zeta_0$, $\zeta\ne -1/6$ on (c) and (d) panels; $\zeta = -1/6$ on (e) and (f) panels. We can see divergences of $p_H$ and $p_h$ at the locations of exponential solutions and $p_H = 0$ 
with $p_h = 0$ at nonstandard singularities. We also confirm $K_3$ regime at large $H$ in all cases presented in Fig.~\ref{D2_4}.

\begin{figure}
\includegraphics[width=1.0\textwidth, angle=0]{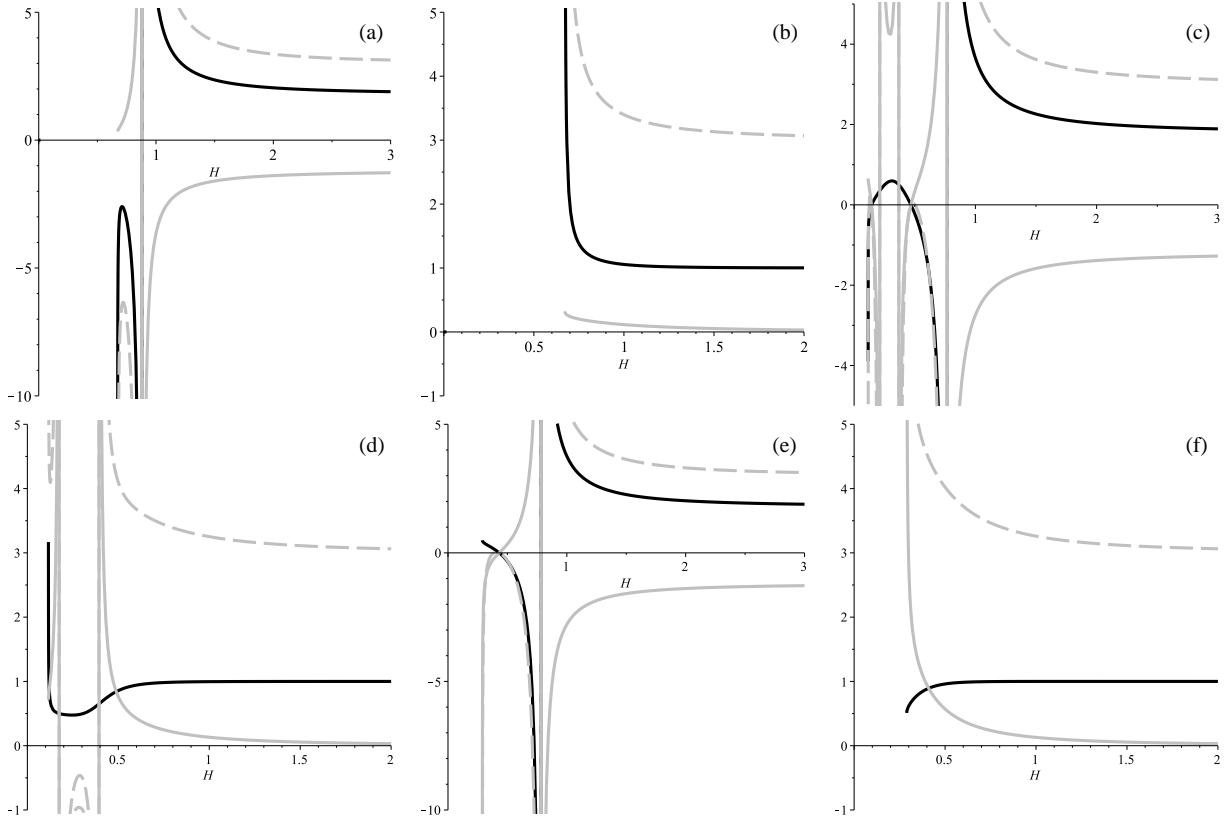}
\caption{Behavior of Kasner exponents in $D=2$ model for $\alpha > 0$ and: $\zeta < \zeta_0$ on (a) and (b) panels; $0 > \zeta > \zeta_0$, $\zeta\ne -1/6$ on (c) and (d) panels; $\zeta = -1/6$ on (e) and (f) panels
(see the text for more details).}\label{D2_4}
\end{figure}

The cases presented in Fig.~\ref{D2_5} and further are more interesting. Indeed, starting from $\zeta > 0$ the domain of definition for $\dot H$ and $\dot h$ cover entire $H>0$ in $\alpha > 0$ case, so that
now we have $H\to 0$ solution. Taking appropriate limits, we can find limits for both $H\to 0$ and $H\to\infty$; we present them in Table~\ref{D.2a}.

\begin{table}[h]
\begin{center}
\caption{Summary of $D=2$ $\Lambda$-term power-law regimes.}
\label{D.2a}
  \begin{tabular}{|c|c|c|c|}
    \hline
     $p_i$ & $H\to 0$ & \multicolumn{2}{c|}{$H\to\infty$}  \\
    \hline
 & both branches & $h_+$ & $h_-$ \\ \hline
$p_H$ & 0 & $ -\frac{9\alpha}{2\sqrt{\alpha^2} - 7\alpha}$ & $\frac{9\alpha}{2\sqrt{\alpha^2}+7\alpha}$ \\ \hline
$p_h$ & $-\frac{2}{3}\alpha\Lambda + \frac{2}{3}$ & $- \frac{3\sqrt{\alpha^2} + 3\alpha}{5\alpha}$ & $\frac{3\sqrt{\alpha^2} - 3\alpha}{5\alpha}$ \\ \hline
$\sum p$ & $-\frac{4}{3}\alpha\Lambda + \frac{4}{3}$ & 3 & 3 \\
    \hline
  \end{tabular}
\end{center}
\end{table}

One can clearly see that both high-energy regimes are Gauss-Bonnet Kasner with $\sum p = 3$. But low-energy regimes are not Kasner, as they do not have $\sum p = 1$. With $p_H = 0$ one can immediately confirm
that $\sum p_i p_j p_k = 0$ for this case which makes it ``generalized Milne'' (see e.g.~\cite{prd09}). Recently it was demonstrated that in vacuum case this regime is forbidden~\cite{PT}, and our analysis
of the vacuum case~\cite{my16a} confirms this.
Let us also note that in all previous numerical studies~\cite{mpla09, prd10, grg10, KPT} we never detected this regime so
this is the first time we see it. But also, this regime is not asymptotic (similar to the Kasner-like regimes in $D=1$), so it is not really reached and the final asymptotic regime is either exponential 
with $H<0$ or nonstandard singularity with $H<0$, depending on the case.

Apart from these regimes, which could be clearly seen from, say, Figs.~\ref{D2_5}(c, d), we can observe usual components like divergences of $p_H$ and $p_h$ at the locations of the exponential solutions as 
well as zeros of $p_H$ and $p_h$ at the locations of nonstandard singularities. Figures~\ref{D2_5}(f, g) illustrate the situation with multiple exponential solutions. We skip the detailed description of all
cases but confirm that the regimes correspond to the description in $\{\dot H, \dot h\}$.

In Fig.~\ref{D2_7}, which corresponds to $\alpha < 0$ case, we once again can see incomplete domain of definition: on (a) and (b) panels it is $H>H_0$ while on (e) and (f) panels it is
$H\in [0; H_{0(1)}) \cup (H_{0(2)}; + \infty)$ -- all according to the description in $\{\dot H, \dot h\}$.

\begin{figure}
\includegraphics[width=1.0\textwidth, angle=0]{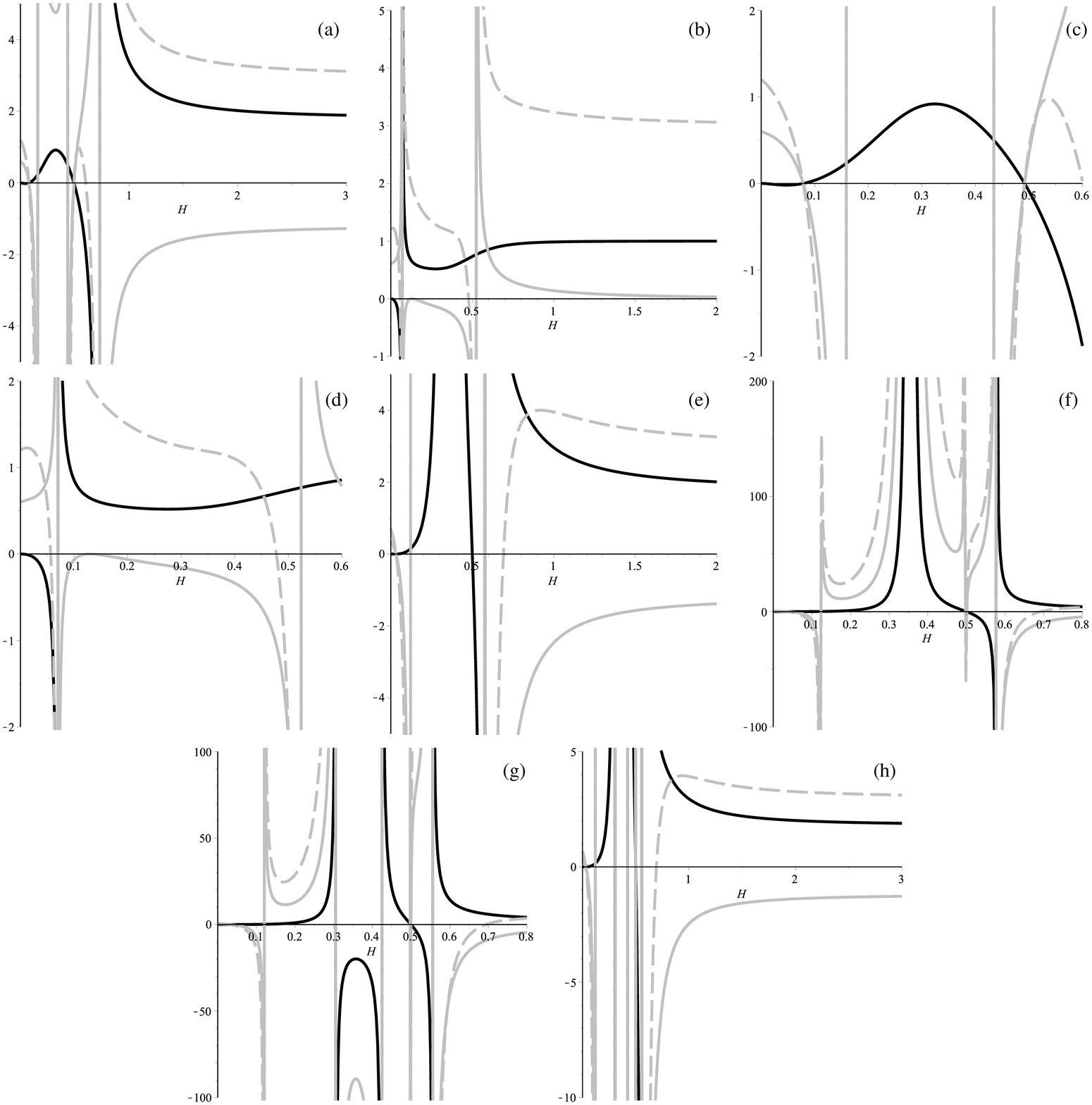}
\caption{Continuation of Fig.~\ref{D2_4} for $\alpha > 0$ and: $0 < \zeta < 15/32$ on (a)--(d) panels; $\zeta = 15/32$ on (e) and (f) panels; $1/2 > \zeta > 15/32$ on (g) and (h) panels
(see the text for more details).}\label{D2_5}
\end{figure}

\begin{figure}
\includegraphics[width=1.0\textwidth, angle=0]{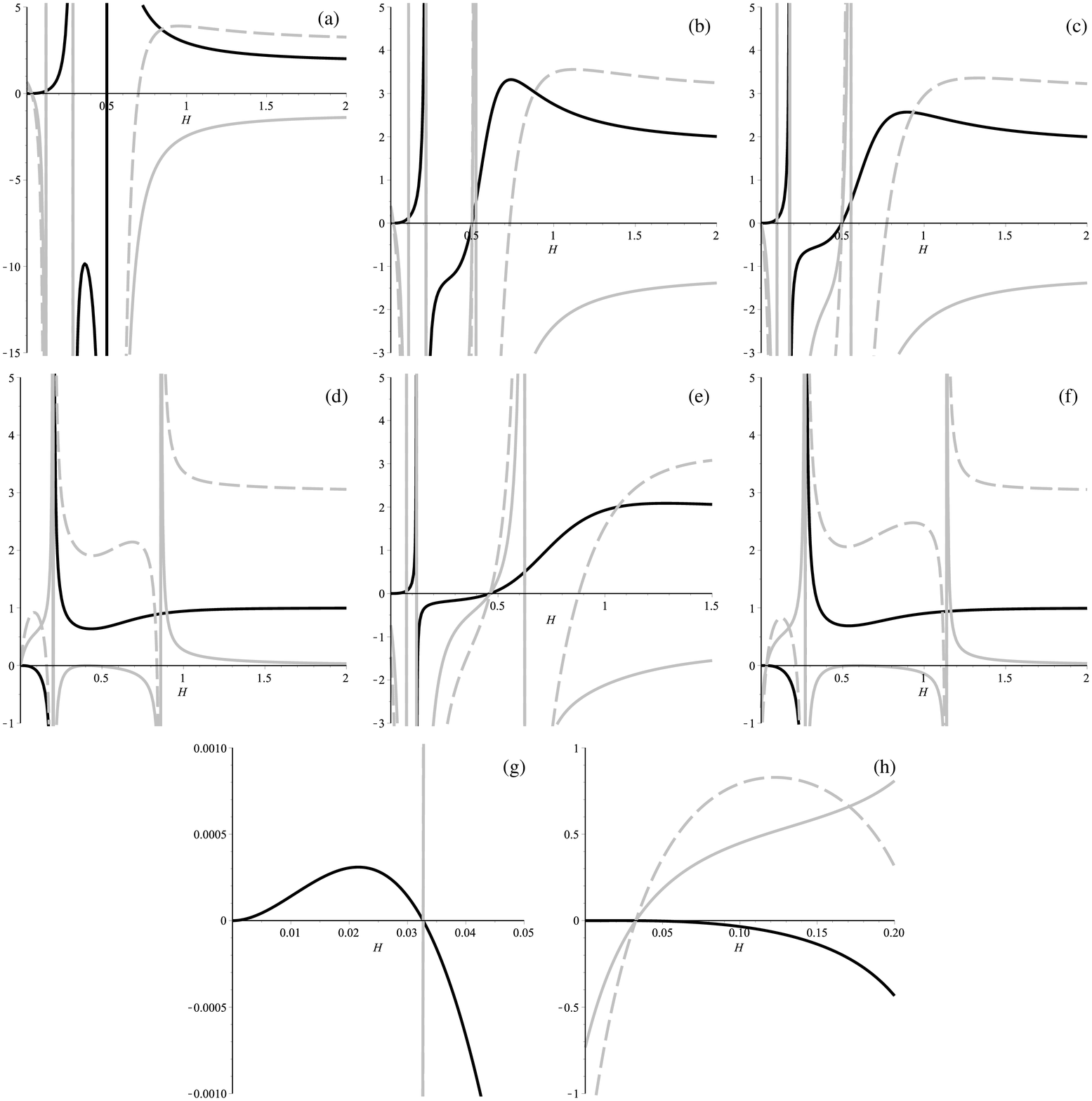}
\caption{Continuation of Figs.~\ref{D2_4} and \ref{D2_5} for $\alpha > 0$ and: $\zeta = 1/2$ on (a) panel; $1 > \zeta > 1/2$ on (b) panel; $\zeta = 1$ on (c) and (d) panels;
$\zeta > 1$ on (e)--(h) panels
(see the text for more details).}\label{D2_6}
\end{figure}

\begin{figure}
\includegraphics[width=1.0\textwidth, angle=0]{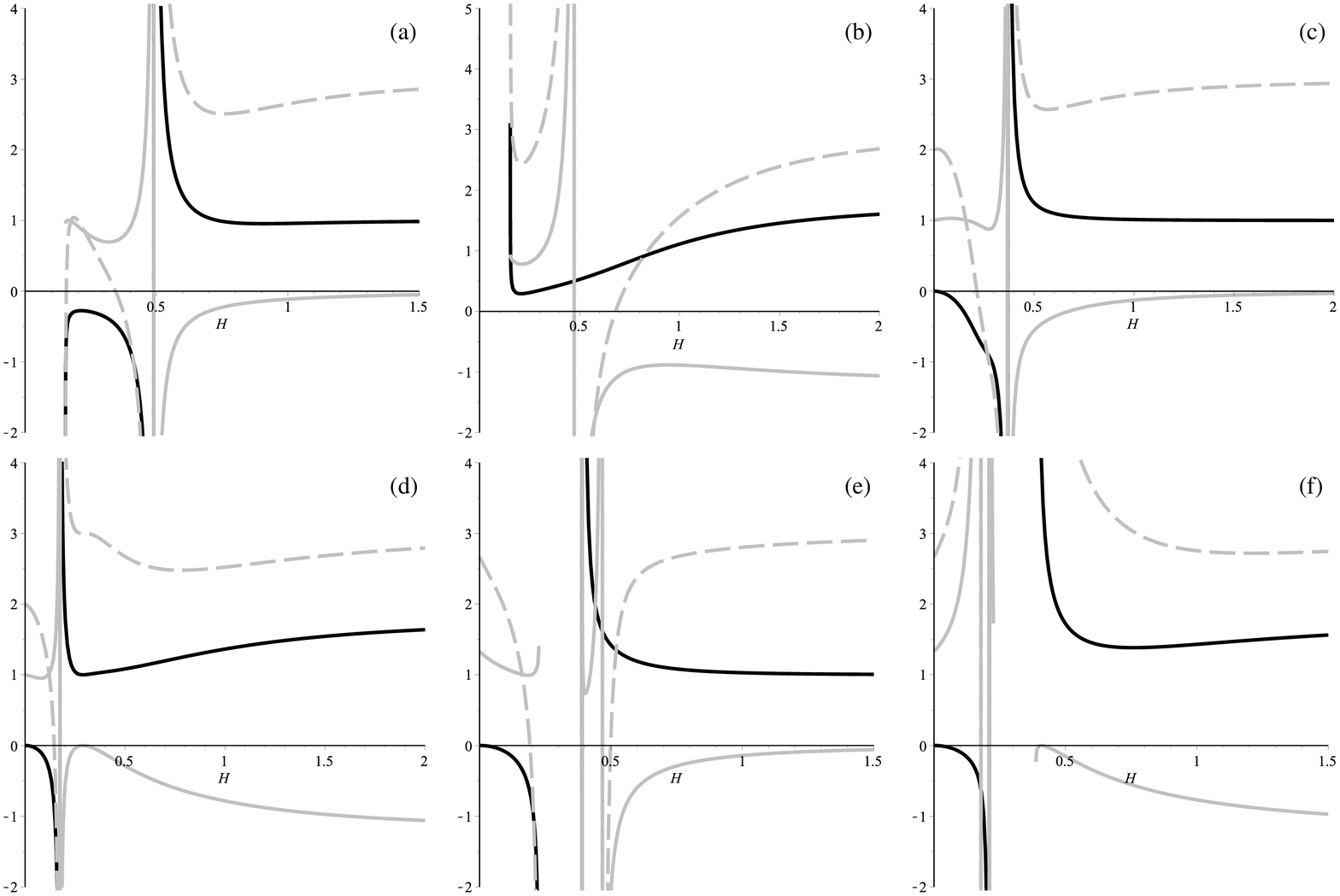}
\caption{Continuation of Figs.~\ref{D2_4}--\ref{D2_6} for $\alpha < 0$ and: $\zeta > 0$ on (a) and (b) panels; $\zeta > -5/6$ on (c) and (d) panels; $\zeta < -5/6$ on (e) and (f) panels
$\zeta > 1$ on (e)--(h) panels
(see the text for more details).}\label{D2_7}
\end{figure}

Finally it is time to summarize all our regimes found. Due to enormous amount of different regimes we spread them into three tables -- \ref{D.2a}, \ref{D.2b} and \ref{D.2c}. In all tables we assume
following notations: $H(\xi_i)$ is the value for Hubble parameter $H$ derived from given value for $\xi_i$ and for given value of $\alpha$ from $\xi_i = \alpha H^2$ relation. Individual $\xi_i$ are
described above and here we just recollect them: $\xi_\pm$ are roots of the quadratic equation from (\ref{D2_exp}), $\xi_{1,\, 2,\, 3}$ are roots of the cubic equation from (\ref{D2_exp}), $\xi_{4,\, 5}$
are roots of the denominator $Q_\pm$ from (\ref{D2_dHdh_2}). Positivity of the discriminant (\ref{D2_hh_discr}) is guaranteed by $H > H_0$ in case when there is only one positive root (like grey curve
in Fig. \ref{D2_0}(a)) and by $H < H_{0(1)}$ combined with $H > H_{0(2)}$ for the case when there are two roots (like black curve in Fig. \ref{D2_0}(b)). Below we briefly comment on some regimes for
each particular table.

In Table \ref{D.2b} we summarized ($\alpha > 0$, $\Lambda < 0$) regimes. One can see that the only exponential solutions in that case are isotropic ones so they do not correspond to any physical cases.

\begin{table}
\begin{center}
\caption{Summary of $D=2$  ($\alpha > 0$, $\Lambda < 0$) $\Lambda$-term regimes.}
\label{D.2b}
  \begin{tabular}{|c|c|c|c|}
    \hline
      Branch & \multicolumn{2}{c|}{Additional conditions} & Regimes  \\
    \hline
 \multirow{12}{*}{$h_+$} & \multirow{3}{*}{$\alpha\Lambda \lessapprox -0.17$} & $H < H_0$ & no solutions \\ \cline{3-4}
                                                                                         &                                                & $H(\xi_-) > H > H_0$ & $nS \to E_{iso}$ \\ \cline{3-4}
                                                                                         &                                                & $H > H(\xi_-)$ & $K_3 \to E_{iso}$ \\ \cline{2-4}
                                                                                        & \multirow{5}{*}{$\alpha\Lambda \gtrapprox -0.17,\, \ne -1/6$} & $H < H_0$ & no solutions \\ \cline{3-4}
                                                                                         &                                                & $H(\xi_4) > H > H_0$ & $nS \to nS$ \\ \cline{3-4}
                                                                                         &                                                & $H(\xi_5) > H > H(\xi_4)$ & $nS \to nS$ \\ \cline{3-4}
                                                                                         &                                                & $H(\xi_-) > H > H(\xi_5)$ & $nS \to E_{iso}$ \\ \cline{3-4}
                                                                                         &                                                & $H > H(\xi_-)$ & $K_3 \to E_{iso}$ \\ \cline{2-4}
                                                                                         &  \multirow{4}{*}{$\alpha\Lambda = -1/6$} & $H < H_0$ & no solutions \\ \cline{3-4}
                                                                                         &                                                & $H(\xi_{4,\,5}) > H > H_0$ & $nS \to nS$ \\ \cline{3-4}
                                                                                        &                                                & $H(\xi_-) > H > H(\xi_{4,\,5})$ & $nS \to E_{iso}$ \\ \cline{3-4}
                                                                                         &                                                & $H > H(\xi_-)$ & $K_3 \to E_{iso}$ \\ \cline{1-4}
                                                                 \multirow{2}{*}{$h_-$}  & \multicolumn{2}{c|}{$ H < H_0$}                & no solutions \\ \cline{2-4}
                                                                                         & \multicolumn{2}{c|}{$ H > H_0$}                & $K_3 \to nS$ \\
    \hline
  \end{tabular}
\end{center}
\end{table}

In Table \ref{D.2c} we summarized ($\alpha > 0$, $\Lambda > 0$) regimes. As we can see from Figs. \ref{D2_1} and \ref{D2_2}, this case is most abundant with different regimes. We can detect up to three anisotropic
exponential solutions for a single $h_+$ branch, plus isotropic solution on $h_-$ branch for the same values of $\alpha$ and $\Lambda$. Also in this case we have $K_3 \to E$ transition -- smooth transition
from Gauss-Bonnet Kasner to anisotropic exponential solution -- the only physically viable regime we detected. Let us also note that this regime exist only for $\alpha\Lambda \leqslant 1/2$.

\begin{table}
\footnotesize
\centering
\caption{Summary of $D=2$ ($\alpha > 0$, $\Lambda > 0$) $\Lambda$-term regimes.}
\label{D.2c}
  \begin{tabular}{|c|c|c|c|}
    \hline
     Branch & \multicolumn{2}{c|}{Additional conditions} & Regimes  \\
    \hline
\multirow{29}{*}{$h_+$} & \multirow{4}{*}{$\alpha\Lambda < 15/32$} & $H < H(\xi_4)$ & $nS^{(-)} \to nS$ \\  \cline{3-4}
 & & $H(\xi_5) > H > H(\xi_4)$ & $nS \to nS$ \\ \cline{3-4}
 & & $H(\xi_1) > H > H(\xi_5)$ & $nS \to E_1$ \\ \cline{3-4}
 & & $H > H(\xi_1)$ & $K_3 \to E_1$ \\ \cline{2-4}
 & \multirow{5}{*}{$\alpha\Lambda = 15/32$} & $H < H(\xi_4)$ & $nS^{(-)} \to nS$ \\  \cline{3-4}
 & & $H(\xi_{2,\,3}) > H > H(\xi_4)$ & $E_{2,\,3} \to nS$ \\ \cline{3-4}
 & & $H(\xi_5) > H > H(\xi_{2,\,3})$ & $nS \to E_{2,\,3}$ \\ \cline{3-4}
 & & $H(\xi_1) > H > H(\xi_5)$ & $nS \to E_1$ \\ \cline{3-4}
 & & $H > H(\xi_1)$ & $K_3 \to E_1$ \\ \cline{2-4}
 & \multirow{6}{*}{$1/2 > \alpha\Lambda > 15/32$} & $H < H(\xi_4)$ & $nS^{(-)} \to nS$ \\ \cline{3-4}
 & & $H(\xi_2) > H > H(\xi_4)$ & $E_2 \to nS$ \\  \cline{3-4}
 & & $H(\xi_3) > H > H(\xi_2)$ & $E_2 \to E_3$ \\ \cline{3-4}
 & & $H(\xi_5) > H > H(\xi_3)$ & $nS \to E_3$ \\  \cline{3-4}
 & & $H(\xi_1) > H > H(\xi_5)$ & $nS \to E_1$ \\  \cline{3-4}
 & & $H > H(\xi_1) $ & $K_3 \to E_1$ \\ \cline{2-4}
 & \multirow{4}{*}{$\alpha\Lambda = 1/2$} & $H < H(\xi_4)$ & $nS^{(-)} \to nS$ \\  \cline{3-4}
 & & $H(\xi_{2,\,3}) > H > H(\xi_4)$ & $E_{2,\,3} \to nS$ \\ \cline{3-4}
 & & $H(\xi_1) > H > H(\xi_{2,\,3})$ & $E_1 \to E_{2,\,3}$ \\ \cline{3-4}
 & & $H > H(\xi_1)$ & $K_3 \to E_1$ \\ \cline{2-4}
 & \multirow{4}{*}{$1 > \alpha\Lambda > 1/2$} & $H < H(\xi_4)$ & $nS^{(-)} \to nS$ \\  \cline{3-4}
 & & $H(\xi_{\{1,\,3\}}) > H > H(\xi_4)$ & $E_{\{1,\,3\}} \to nS$ \\  \cline{3-4}
 & & $H(\xi_5) > H > H(\xi_{\{1,\,3\}})$ & $E_{\{1,\,3\}} \to nS$ \\  \cline{3-4}
 & & $H > H(\xi_5)$ & $K_3 \to nS$ \\  \cline{2-4}
 & \multirow{3}{*}{$\alpha\Lambda = 1$} & $H < H(\xi_1)$ & $E_1 \to nS$ \\ \cline{3-4}
 & & $H(\xi_5) > H > H(\xi_1)$ & $E_1 \to nS$ \\ \cline{3-4}
 & & $H > H(\xi_5)$ & $K_3 \to nS$ \\  \cline{2-4}
 & \multirow{3}{*}{$\alpha\Lambda > 1$} & $H < H(\xi_1)$ & $E_1 \to E_1^{(-)}$ \\ \cline{3-4}
 & & $H(\xi_5) > H > H(\xi_1)$ & $E_1 \to nS$ \\ \cline{3-4}
 & & $H > H(\xi_5)$ & $K_3 \to nS$ \\  \cline{1-4}
\multirow{7}{*}{$h_-$} & \multirow{2}{*}{$\alpha\Lambda < 1$} & $H < H(\xi_+)$ & $E_{iso}^{(-)} \to E_{iso}$ \\ \cline{3-4}
& & $H > H(\xi_+)$ & $K_3 \to E_{iso}$ \\ \cline{2-4}
& \multirow{2}{*}{$\alpha\Lambda = 1$}  & $H < H(\xi_+)$ & $nS \to E_{iso}$ \\ \cline{3-4}
& & $H > H(\xi_+)$ & $K_3 \to E_{iso}$ \\ \cline{2-4}
& \multirow{3}{*}{$\alpha\Lambda > 1$} & $H < H(\xi_4)$ & $nS \to nS^{(-)}$ \\ \cline{3-4}
& & $H(\xi_1)> H > H(\xi_4)$ & $nS \to E_{iso}$ \\ \cline{3-4}
& & $H > H(\xi_1)$ & $K_3 \to E_{iso}$ \\
     \hline
  \end{tabular}
\end{table}

Finally in Table \ref{D.2d} we presented the results for $\alpha < 0$. In there, $E_2$ and $E_3$ exponential solutions require additional clarification -- they are anisotropic but they give $h_\pm/H > 0$,
so that both three-dimensional and extra-dimensional manifolds are expanding. We treat it as violation of viability and so discard them. This situation is partially change for
$\alpha\Lambda \leqslant -3/2$ -- in that case $E_3$ solution (on $h_+$ branch) change its sign to $h_+/H \leqslant 0$ while its $E_2$ counterpart disappears. Also, for exact $\alpha\Lambda = -3/2$ relation we have
$h(t) \to 0$, in which we recover the regime quite similar to ``geometric frustration'' one, described in~\cite{CGP1}. We discuss it further is Discussions section.

\begin{table}
\begin{center}
\caption{Summary of $D=2$ $\alpha < 0$ $\Lambda$-term regimes.}
\label{D.2d}
  \begin{tabular}{|c|c|c|c|c|}
    \hline
 $\Lambda$  &  Branch & \multicolumn{2}{c|}{Additional conditions} & Regimes  \\
    \hline
\multirow{6}{*}{$\Lambda < 0$} & \multirow{4}{*}{$h_+$} & \multicolumn{2}{c|}{$ H < H_0$} & no solutions \\ \cline{3-5}
& & \multicolumn{2}{c|}{$H(\xi_{4,5}) >  H > H_0$} & $nS \to nS$ \\ \cline{3-5}
& & \multicolumn{2}{c|}{$H(\xi_-) > H >  \xi_{4,5})$} & $nS \to E_{iso}$ \\ \cline{3-5}
& & \multicolumn{2}{c|}{$H > H(\xi_-)$} & $K_3 \to E_{iso}$ \\ \cline{2-5}
& \multirow{2}{*}{$h_-$} & \multicolumn{2}{c|}{$ H < H_0$} & no solutions \\ \cline{3-5}
& & \multicolumn{2}{c|}{$ H > H_0$} & $K_3 \to nS$ \\ \cline{1-5}
\multirow{17}{*}{$\Lambda > 0$} & \multirow{8}{*}{$h_+$} & \multirow{3}{*}{$\alpha\Lambda \geqslant -5/6$} & $H < H(\xi_4)$ & $nS^{(-)} \to nS$ \\ \cline{4-5}
& & & $H(\xi_+) > H > H(\xi_4)$ & $nS \to E_{iso}$ \\ \cline{4-5}
& & & $H > H(\xi_+)$ & $K_3 \to E_{iso}$ \\ \cline{3-5}
& & \multirow{5}{*}{$\alpha\Lambda < -5/6$} & $H < H_{\xi_{4,5}}$ & $nS^{(-)} \to nS$ \\ \cline{4-5}
& & & $H_{0(1)} > H > H(\xi_{4,5})$ & $nS \to nS$ \\ \cline{4-5}
& & & $H_{0(2)} > H > H_{0(1)}$ & no solutions \\ \cline{4-5}
& & & $H(\xi_3) > H > H_{0(2)}$ & $nS \to E_3$ \\ \cline{4-5}
& & & $H > H(\xi_3)$ & $K_3 \to E_3$ \\ \cline{2-5}
& \multirow{9}{*}{$h_-$} & \multirow{2}{*}{$\alpha\Lambda \geqslant -5/6$} & $H < H(\xi_-)$ & $E_{iso}^{(-)} \to E_{iso}$ \\ \cline{4-5}
& & & $H > H(\xi_-)$ & $K_3 \to E_{iso}$ \\ \cline{3-5}
& & \multirow{4}{*}{$-3/2 < \alpha\Lambda < -5/6$} & $H < H(\xi_2)$ & $E_{2}^{(-)} \to E_2$ \\ \cline{4-5}
& & & $H_{0(1)} > H > H(\xi_2)$ & $nS \to E_2$ \\ \cline{4-5}
& & & $H_{0(2)}> H > H_{0(1)}$ & no solutions \\ \cline{4-5}
& & & $H > H_{0(2)}$ & $K_3 \to nS$ \\ \cline{3-5}
& & \multirow{3}{*}{$\alpha\Lambda \leqslant -3/2$} & $H < H_{0(1)}$ & $nS^{(-)} \to nS$ \\ \cline{4-5}
& & & $H_{0(2)}> H > H_{0(1)}$ & no solutions \\ \cline{4-5}
& & & $H > H_{0(2)}$ & $K_3 \to nS$ \\ \cline{4-5}
 \hline
  \end{tabular}
\end{center}
\end{table}

This finalize our study of $D=2$ $\Lambda$-term regimes. Unlike $D=1$ case, now we have physically viable regimes $K_3 \to E$. Generally, the dynamics is much more abundant then in both $D=1$ case and
$D=2$ vacuum counterpart~\cite{my16a}. We discuss this, as well as $D=1$ cases, in detail below.

\section{Discussions}

In this paper we have considered Einstein-Gauss-Bonnet cosmological model with $D=1$ and $D=2$ extra dimensions in presence of $\Lambda$-term. In this section we summarize our finding and discuss them.
Before discussing each particular case, let us make several important notes. First one is related to the power-law solutions. We can clearly see that in both cases, unlike
vacuum regimes~\cite{my16a}, we do not have low-energy Kasner regimes. But this is natural -- indeed, in presence of $\Lambda$-term it will eventually start to dominate and turn any low-energy regime into 
the exponential
one. Also, formal consideration of power-law regimes in pure Gauss-Bonnet gravity (see, e.g.,~\cite{prd09, grg10, PT}) forbid power-law regimes to exist in presence of $\Lambda$-term. Indeed,
if we consider, say, constraint equation $\sum H_i H_j H_k H_l = \rho$ in power-law {\it Ansatz} $a_i(t) \propto t^{p_i}$, it takes the form $t^{-4} \sum p_i p_j p_k p_l = \rho$. Now one can see that
in vacuum ($\rho=0$) we can cancel $t^{-4}$ and arrive just to $\sum p_i p_j p_k p_l = 0$ -- well-known condition for the power-law solutions to exist in vacuum Gauss-Bonnet gravity. If $\rho$ is dynamical, say,
perfect fluid, constraint equation also could be solved under additional relation between the equation of state and sum of Kasner exponents~\cite{grg10}. But if $\rho$ is nonzero constant ($\Lambda$-term), constraint
equation cannot be solved for constant $p_i$ and $\Lambda$, which means power-law solutions cannot exist in presence of $\Lambda$-term. Yet, we clearly see them -- at least, $K_3$. This could be explained 
as follows: $\Lambda$-term
is constant while we have dynamical evolution for $H$. And in high-energy regime $H_i \ggg \Lambda$ so that we could consider $\Lambda/H_i \approx 0$ in that case and recover Gauss-Bonnet power-law solutions.

Another surprise with power-law solutions is the existence of the generalized Milne-like solutions, although unstable. In~\cite{PT} we clearly demonstrated that in vacuum,
the generalized Milne solution cannot exist as it leads to degenerative system with unconstrained Hubble parameters. It was decided that this is an artifact caused by the fact that we neglect lower-order
contribution while building power-law solutions. Yet, we detected these solutions, although as an intermediate regime between nonstandard singularities or exponential solutions.

Other important notes regard exponential solutions. First, in $D=2$ case exponential solutions are governed by two equations -- quadratic and cubic. On contrary, in our paper dedicated to the
exponential solutions in lower-dimensional Einstein-Gauss-Bonnet cosmologies~\cite{CPT1}, exponential solutions for this case are reported to be governed by cubic equation alone (one can check that the
cubic equations from this paper and from~\cite{CPT1} are the same). In this paper anisotropic solutions are governed by the roots of this cubic while isotropic -- only by roots of this quadratic, but
potentially in other cases additional roots could give rise to additional exponential solutions, so additional study of exponential solutions in both Einstein-Gauss-Bonnet and more general Lovelock gravity
is required.

Another note which regard exponential solutions is related to their stability. Indeed, stability of exponential solutions was addressed in~\cite{my15} and the results for vacuum case~\cite{my16a} are in
perfect agreement with them. Yet, from Figs. \ref{D2_1} and \ref{D2_2} one can clearly see directional stability or even instability of certain exponential solutions for $15/32 \geqslant \zeta \geqslant 1/2$. We have not
reported any cases of the directional stability in~\cite{my15}, but see them in current research from phase portraits; this also require additional investigation.

As a final point, we want to stress readers' attention to the situation we call ``nonstandard singularity''. As we can see from the equations of motion (\ref{dyn_gen}), they are nonlinear with
respect to the highest derivative\footnote{Actually, this is one of the definitions of Lovelock (and Gauss-Bonnet as its particular case) gravity: it is well-known~\cite{etensor1, etensor2, etensor3} that
the Einstein tensor is, in any dimension, the only symmetric and
conserved tensor depending only on the metric and its first and
second derivatives (with a linear dependence on second
derivatives). If one drops the condition of linear dependence on
second derivatives, one can obtain the most general tensor which
satisfies other mentioned conditions -- Lovelock
tensor~\cite{Lovelock}.}, so formally we can solve them with respect to the highest derivative. Then, the highest derivative is expressed as a ratio of two polynomials, both
depending on $H$. And there could be a situation when the denominator of this expression is equal to zero while numerator is not. In this case $\dot H$ diverges while $H$ is (generally) nonzero and
regular. In our study we saw nonstandard singularities with divergent $\dot h$ or both $\dot h$ and $\dot H$ at nonzero or sometimes zeroth $H$.
This kind of singularity is ``weak'' by Tipler's classification~\cite{Tipler}, and ``type II''
in classification by Kitaura and Wheeler~\cite{KW1, KW2}. Recent studies of the singularities of this kind in the cosmological context in Lovelock and Einstein-Gauss-Bonnet gravity
demonstrates~\cite{CGP2, mpla09, grg10, KPT, prd10} that their presence is not suppressed and they are abundant for a wide range of initial conditions and parameters and sometimes~\cite{prd10} they are the
only option for future behavior.

With these points noted, let us turn to summarizing particular cases. First of them, $D=1$ case, have GB Kasner $K_3$ as the only high-energy regime and singular Kasner-like $\tilde K_1^S$ regime as
the only low-energy regime. This makes the difference between this and the vacuum cases -- the latter have non-singular low-energy Kasner regime and so have smooth transition between high- and low-energy
regimes. The $\Lambda$-term $D=1$ case lacks this transition so we cannot restore Friedmann-like behavior in this case. Intermediate-energy regimes include nonstandard singularities and exponential
regimes, and the latter are presented only by isotropic regimes. This is expected -- indeed, in~\cite{CPT1} we demonstrated that for $\Lambda$-term $D=1$ model there are two possible exact exponential
solutions -- isotropic solution and anisotropic one, but the latter with $h \in \mathds{R}$. The fact that $h$ is unconstrained feels unphysical and later in~\cite{my15} it was demonstrated that anisotropic
solution is marginally/neutrally stable. Finally in this study we clearly demonstrate that this solution formally exist but is never reached. So that it is natural to extend this conclusion on other
exponential solutions from~\cite{CPT1, CPT3} with one or more unconstrained Hubble parameters -- they are unphysical and cannot be reached which leaves us only with solutions proven to be stable in~\cite{my15}.
What was unexpected is the presence of two distinct isotropic exponential solutions in $D=1$ case -- it cannot be seen from the study of exact exponential solutions~\cite{CPT1}, but here we detect them.
This fact also indicate that additional study of exponential solutions is required. So that in $D=1$ case we have no viable late-time regimes -- the only nonsingular regime is isotropic exponential
expansion and it is definitely not what we observe.

Another case considered is $D=2$. It is very different from $D=1$ both in power-law and exponential solutions. The former of them have GB Kasner $K_3$, just like $D=1$, but similarities end here. Unlike
$D=1$ case, $D=2$ one does not have stable low-energy regime. Indeed, we saw that $H=0$ point in $D=2$ is regular point of dynamical evolution and so the dynamics is prolonged to $H<0$ domain. This is
very different from what we saw in vacuum~\cite{my16a} and $D=1$ $\Lambda$-term cases. Indeed, checking $h(H)$ and $H(h)$ expressions and curves in vacuum case~\cite{my16a} clearly demonstrate that
there are only two options -- either $H\to 0$ with $h\to 0$, which gives us regular low-energy Kasner regime $K_1$, or $H\to 0$ with $h\to\pm\infty$, which gives us singular low-energy Kasner
regime\footnote{Singular low-energy Kasner regime arises from negative Kasner exponent $p_i$ -- indeed, in case $p_i < 0$ we have $a_i(t) \propto 1/t^{\alpha_i}$ with $\alpha_i = - p_i > 0$, making
$a_i(t)$ divergent at $t=0$.}. On contrary, from (\ref{D2_hh}) and Figs.~\ref{D2_0}(e)--(i) one can see that in $D=2$ case for $H\to 0$ we have regular and nonzero $h$. This results in absence of
regular low-energy (power-law) regime and in another feature, which we discuss a bit later.

So that the regimes which reach $H=0$ continue to $H<0$ domain and, due to the $H\to -H$ symmetry of the equations of motion (\ref{D2_dHdh_1})--(\ref{D2_ch_dh}), face the regime which is dual to the closest to $H=0$.
Since the only regimes that could be presented are nonstandard singularity or exponential solution, possible transitions include pairwise evolution between such singularities and expanding/contracting exponential solutions. In reality
we have contracting isotropic exponential solution anisotropically bounce to expanding isotropic exponential solution (see Fig.~\ref{D2_1}(g) and Fig.~\ref{D2_3}(d)) as well as anisotropic solution with expanding
three-dimensional space transits to regime with contracting three-dimensional space (see Fig.~\ref{D2_2}(h) and the opposite -- contracting transits to expanding (see Fig.~\ref{D2_3}(f, h).

For the regimes which do not reach $H=0$, future evolution is represented by either nonstandard singularities or exponential solutions. The former of them cannot correspond to a viable
regime but some of the latter can. In $D=1$ case we have only isotropic solutions but $D=2$ abundant with anisotropic ones as well. For $\alpha > 0$, $\Lambda < 0$ there are only isotropic solutions (see
Table~\ref{D.2b}) but in $\alpha > 0$, $\Lambda > 0$  case there are up to three different anisotropic solutions. This number comes from the number of possible roots of bicubic equation and is in
agreement with~\cite{CPT1}. Let us note that in this case anisotropic solutions exist only for $\alpha \Lambda \leqslant 1/2$ and only in $h_+$ branch (see Table~\ref{D.2c}); $h_-$ branch has isotropic
solutions only. Finally for $\alpha < 0$, as we can see from Table~\ref{D.2d}, for $\Lambda < 0$ we have only isotropic solutions for $h_+$ and no exponential solutions for $h_-$. For $\Lambda > 0$
situation changes a bit: for both branches we have isotropic solution for $\alpha\Lambda \geqslant -5/6$ and anisotropic for $\alpha\Lambda < -5/6$. One of these two anisotropic solutions ($E_2$ on $h_-$
branch) always have $h(t)>0$ while another ($E_3$ on $h_+$ branch) -- only as long as $\alpha\Lambda > -3/2$, because at $\alpha\Lambda < -3/2$ it has $h(t) < 0$ and so contraction of extra dimensions is restored.
Also, from Table~\ref{D.2d} one can see that only $E_3$ is the ``endpoint'' for $K_3 \to E_3$ transition while $E_2$ does not have high-energy regime as a past asymptote.
At $\alpha\Lambda = -3/2$, $E_2$ solution disappears while
$E_3$ solution has $h(t) \to 0$ so that extra dimensions ``stabilize''
-- their ``size'' (in terms of the scale factor) reaches some constant value. This is very similar to the stabilization of the extra dimensions size due to the ``geometric frustration'', described in~\cite{CGP1}
and further analyzed in~\cite{CGP2, CGPT}. But these two cases have a huge difference -- ``geometric frustration'' case have negative spatial curvature of extra dimensions and special range of couplings while
this case is spatially flat and we have exact relation between the couplings.
The regime with constant-size extra dimensions is of additional importance -- in case if extra dimensions are topologically compact, the total action could be expressed as four-dimensional action plus some corrections -- the similar
way as it is done for Kaluza-Klein theory. If it is done, one could pose additional constraints on the theory from accelerator physics.

Let us note that we have not seen solutions of this type (with $h(t) \to 0$) neither in vacuum case~\cite{my16a} nor in $D=1$ case -- it is another feature of $h(H)$ relation in $D=2$ case, discussed above.
Indeed, if we substitute $h=0$ into the general constraint equation (\ref{con2_gen}), it takes the form $6H^2 = \Lambda$ -- so that in vacuum case it is always $H=0$ -- either $K_1$ or nonstandard singularity.
If we further substitute $h=0$ and $6H^2 = \Lambda$ into one of the dynamical equations (\ref{H_gen}), it takes the form $\dot H = 0$ - in both vacuum and $\Lambda$-term cases. In the former of them
we additionally have $H=0$ which corresponds to the low-energy Kasner regime while in the latter we have $H\ne 0$ which corresponds to the exponential solution. 
But this scheme does not worked for $D=1$ $\Lambda$-term case due to degeneracy between $H$ and $h$ -- it is lowest-dimensional case and in higher dimensions $D \geqslant 2$ this degeneracy is removed.

\section{Conclusions}

To conclude, we performed thorough analysis of $\Lambda$-term regimes in Einstein-Gauss-Bonnet gravity in two lowest number of dimensions -- five and six. We have considered the manifold which is a product of
three-dimensional (which represents our Universe) and extra-dimensional (in our case with $D=1,\,2$ dimensions) parts. This separation is quite natural as with it we could describe natural compactification.
Our analysis demonstrate that generally $\Lambda$-term models have much mode abundant dynamics then vacuum cases~\cite{my16a}. Our investigation also suggest that in $D=1$ model there are no physically viable
 regimes. On contrary, $D=2$ case have smooth transitions from high-energy Kasner regime to anisotropic exponential solutions with contracting
 extra dimensions. In one particular case $\alpha\Lambda = -3/2$ with $\alpha < 0$ and $\Lambda > 0$ the size of extra dimensions (in the sense of the scale factor) reaches constant value (and so the 
 expansion rate $h(t)\to 0$), making this case similar to
 the regime described in spatially-curved ``geometric frustration'' model~\cite{CGP1}.

 Both $D=1$ and $D=2$ cases lack regular low-energy regime -- in the former of them it is singular (one faces finite-time singularity while reaching low-energy Kasner regime) and for the latter $H=0$ is not an endpoint
 and the evolution continues to $H<0$ domain until either nonstandard singularity or exponential solution is reached. So that in $D=2$ case we have interesting regimes like the transition from isotropic exponential contracting to
 isotropic exponential expansion (like a bounce) and anisotropic regimes with contracting three-dimensional spaces turn to expansion and vice versa.

Lack of low-energy regimes in $D=2$ as well as presence of $h=0$ anisotropic exponential solution have the same cause -- $h(H)$ expression in $D=2$ case is distinct from both $D=1$ and the entire vacuum case~\cite{my16a}. Indeed,
in both $D=1$ and the vacuum case, $h(H)$ and $H(t)$ curves have either $h\to 0$ or $h\to\pm\infty$ as $H\to 0$. In the former case we have low-energy nonsingular Kasner regime, in the latter -- the same but singular. Also one can see
that we cannot have $h=0$ while $H\ne 0$, which prevent corresponding exponential solutions to exist. On contrary, one can see that in $D=2$ we have $h\ne 0$ at $H=0$ so that $h=0$, $H\ne 0$ exponential
solutions exist while low-energy regimes are absent. With the same argumentation, $h=0$, $H\ne 0$ exponential solutions could exist in higher-order Lovelock models as well, except for lowest possible
dimensions, like $D=3$ for cubic Lovelock, $D=4$ for quadric and so on.

Overall, present study brought us several unexpected results -- for power-law solutions, we found singular low-energy Kasner-like  behavior for $D=1$ and Milne-like behavior for $D=2$. Both regimes are supposed to
be forbidden to exist in presence of Lambda-term, yet the analysis in term of Kasner exponents points on them. Of course, none of these regimes are reached, but the fact that analysis points on them could indicate that
they formally could exist but are unstable -- so this situation is in need for the additional investigation. Exponential solutions also behave not exactly as expected -- multiple distinct isotropic solutions for both $D=1$
and $D=2$ as well as directional stability of anisotropic solutions in $D=2$ clearly indicate need of additional study of exponential solutions as well.

Low-dimensional $\Lambda$-term case demonstrated interesting dynamics for both $D=1$ and $D=2$ with some unexpected features. In forthcoming paper we consider $D=3$ and generic $D \geqslant 4$ $\Lambda$-term cases
and finalize our study $\Lambda$-term case.

\begin{acknowledgments}
This work was supported by FAPEMA under project BPV-00038/16.
\end{acknowledgments}

\end{document}